\documentclass[12pt,aps,pra,reprint,showpacs]{revtex4-1}   

\usepackage{amsmath}    
\usepackage{graphicx}   
\usepackage{rotating}
\usepackage{lipsum}

\begin{document}

\title{An item response theory evaluation of the Light and Spectroscopy Concept Inventory national data set}

\author{Colin S.\ Wallace}
\affiliation{Department of Physics and Astronomy, University of North Carolina at Chapel Hill, Chapel Hill, North Carolina 27599} 
\email{cswallace@email.unc.edu} 

\author{Timothy G.\ Chambers}
\affiliation{Department of Materials Science and Engineering, University of Michigan, Ann Arbor, Michigan 48109}
\email{timchamb@umich.edu}

\author{Edward E.\ Prather}
\affiliation{Center for Astronomy Education (CAE), Department of Astronomy, Steward Observatory, University of Arizona, Tucson, Arizona 85721}
\email{eprather@as.arizona.edu} 

\date{\today}

\begin{abstract}

This paper presents the first item response theory (IRT) analysis of the national data set on introductory, general education, college-level astronomy teaching using the Light and Spectroscopy Concept Inventory (LSCI).  We used the difference between students'
pre- and post-instruction IRT-estimated abilities as a measure of learning gain.  This analysis provides deeper insights than prior publications into both the LSCI as an instrument and into the effectiveness of teaching and learning in introductory astronomy courses.  Our IRT analysis supports 
the classical test theory findings of prior studies using the LSCI with this population.  In particular, we found that  students in classes that used active learning strategies at least 25\% of the time had 
average IRT-estimated learning gains that were approximately 1 logit larger than students in classes that spent less time on active learning 
strategies.  We also found that instructors who want their classes to achieve an improvement in abilities of average $\Delta \theta = 1$ logit must spend at 
least 25\% of class time on active learning strategies.  However, our analysis also powerfully illustrates the lack of insight into student learning that is revealed by looking at a single measure of learning gain, such as average $\Delta \theta$.  Educators and researchers should also examine the distributions of students' abilities pre- and post-instruction in
order to understand how many students actually achieved an improvement in their abilities and whether or not a majority of students have moved to post-abilities significantly greater than the national average.

\end{abstract}

\maketitle

\section{Introduction}
\label{intro}
We report on our item response theory (IRT) analysis of a national data set of 3205 students' matched pre- and post-responses to the Light and Spectroscopy Concept Inventory (LSCI) \cite{bardar07,LSCI}.  The LSCI is a twenty-six item multiple-choice assessment instrument designed to measure students' conceptual understandings and reasoning abilities on topics involving the properties of light, the luminosity-area-temperature relationship, Wien's law, the Doppler shift, and spectroscopy.  All students in the data set were enrolled in one of sixty-nine different introductory, general education, college-level astronomy classes (hereafter, Astro 101) from across the United States (with one class in Ireland), representing twenty-nine colleges and universities, including associate (2-year) colleges, baccalaureate colleges (4-year primarily bachelor-granting institutions), master's colleges and universities (4-year primarily masters- and bachelors-granting universities), and doctorate-granting (research) universities.  Class sizes ranged from less than ten to over 400 students. 
In a previous publication, students' responses from this national data set were used to investigate the relationship between interactive teaching, classes' learning gains, and class size and institution type \cite{prather09}.  Subsequent studies examined how interactive instruction and students' ascribed (e.g., race) and achieved characteristics (e.g., college grade point average) are related to students' learning \cite{rudolph10}, and how classical test theory (CTT) statistics and individual students' performances change pre- to post-instruction \cite{schlingman12}.

IRT has a number of potential advantages over CTT with respect to the analysis of concept inventory data.  CTT statistics do not estimate the underlying abilities of students independent of the items to which they responded \cite{hambleton93}.  In contrast, when the assumptions of IRT hold and the model fits the data, an IRT analysis can estimate students' abilities and item properties independent of one another \cite{rupp06}.  IRT models have been used in an increasing number of physics and astronomy education investigations, including analyses of the Force Concept Inventory \cite{han15, planinic10, wang10}, the Mechanics Baseline Test \cite{cardamone12}, the Conceptual Survey of Electricity and Magnetism \cite{planinic06}, the Star Properties Concept Inventory \cite{wallace10}, the Newtonian Gravity Concept Inventory \cite{williamson13}, and the Astronomy and Space Science Concept Inventory \cite{sadler10}.  

We performed an IRT analysis of the LSCI national data set in order to move beyond the limitations of CTT, gain further insights into the functioning of the 
LSCI's items, test the robustness of our earlier analyses of the LSCI national data set, and investigate the capacity of active engagement instruction to evolve individual students' underlying astronomy reasoning abilities.  This paper is organized as follows.  In Section \ref{data}, we demonstrate that our data set satisfies the assumptions of IRT and that the model fits the data.  Section \ref{results} presents the results of our analysis.  Our conclusions are in Section \ref{conclusions}.

\section{Data Analysis}
\label{data}

\subsection{Selecting an IRT model}
\label{selecting}
For our analysis, we initially attempted to fit both a two parameter logistic (2PL) and a three parameter logistic (3PL) model to the data.  The 2PL model can be 
written as 
\begin{equation}
\label{2pl}
P(X_{pi} = 1|\theta_{p}, a_{i}, b_{i}) = \frac{\exp [a_{i} (\theta_{p} - b_{i})]}{1 + \exp [a_{i} (\theta_{p} - b_{i})]} ~,
\end{equation}
where $P(X_{pi} = 1)$ represents the probability that a student $p$ of ability $\theta_p$ correctly answers an item $i$ with difficulty $b_i$ and discrimination $a_i$. 
The 3PL model is similar to the 2PL model, except the former includes a third item parameter, $c_i$, which is called the guessing parameter.  This guessing parameter takes into account the fact that there may be some items for which students with extremely low abilities $\theta_p$ still have a nonzero probability of 
giving the correct response.  The 3PL model can be written as
\begin{equation}
\label{3pl}
P(X_{pi} = 1|\theta_{p}, a_{i}, b_{i}, c_{i}) = c_{i} + (1-c_{i}) \frac{\exp [a_{i} (\theta_{p} - b_{i})]}{1 + \exp [a_{i} (\theta_{p} - b_{i})]} ~.
\end{equation}
Readers looking for a pedagogical treatment of these IRT models should consult Embretson  and Reise \cite{embretson00}, Hambleton and Jones 
\cite{hambleton93}, Harris \cite{harris89}, or Wallace and Bailey \cite{wallace10}, and references therein.

We used the IRTPRO software \cite{cai2011} to estimate item parameters and student abilities.  We selected the MML estimation procedure for estimating item
parameters and the EAP estimation procedure for estimating students' abilities; see Baker and Kim \cite{baker04} for details on these estimation procedures.  It is very important to note that the 
logit scale was anchored such that the mean ability of the post-instruction scores is 0 logits.

We first tried to fit the 2PL and 3PL models to both the pre- and post-instruction responses of all 3205 students to all twenty-six of the LSCI's items.  However, 
when we calibrated the items to the pre-instruction responses, we got quite different results from when we calibrated the items to the post-instruction responses.  
Furthermore, all our attempts to fit 2PL and 3PL models to the pre-instruction data consistently yielded poor goodness-of-fit statistics.  This result makes sense.
We previously found that the average pre-instruction scores for classes were clustered in the very narrow range of 24\% $\pm$ 2\% \cite{prather09}.  When we
look at individual students' pre-instruction scores, we find that 57\% of students score at or below 25\% correct, which is the most probable score one would expect to receive if 
one is purely guessing \cite{schlingman12}.  This strongly suggests that, pre-instruction, many students posses very little of the latent trait measured by the LSCI, 
which severely limits the utility of the pre-instruction data for producing
accurate estimates of the item parameters.  Consequently, we used only students' post-instruction responses to estimate the item parameters.  We then used these established item parameter values when we estimated students' pre-instruction abilities.

We found that the $\chi^2$ goodness-of-fit statistics for many individual items on the LSCI were significantly better for
the 3PL model than the 2PL model.   While adding another free parameter ($c_i$) will almost always improve model fit, the degree to which the fit improved was 
greater than one would expect from simply adding a free parameter.  This can be seen by calculating the root mean square error of approximation (RMSEA) \cite{maydeu10}
for both the 2PL and 3PL models.  The RMSEA is a statistic that measures the goodness-of-fit of a model relative to the number of parameters in the model such that 
merely adding a new parameter cannot reduce the value of the RMSEA unless the new parameter actually models a real feature present in the data (e.g., a lower 
asymptote in the probability of a correct response \emph{{\`a} la} the 3PL model).  The 2PL model's RMSEA is 0.07 while the 3PL model's RMSEA is 0.06.  
This suggests that guessing was a significant factor in many students'
responses on many items.  While some items ended up with values of $c_i$ near zero (suggesting that these items had many powerful distractors), other items saw 
as many as 40\% of low-ability students answer correctly.  Items with high guessing parameters tended to be those with only one or two frequently chosen distractors.  In a 
previous publication, we found that these one or two distractors, plus the correct answer, tend to dominate the answer choices actually selected by students, which 
implies that these distractors are well-matched to students' conceptual and reasoning difficulties \cite{schlingman12}.  Because guessing appears to be an important
component of students' responses, we abandoned the 2PL model.  All results reported in the rest of this paper were obtained using the 3PL model.

\begin{figure*}
\includegraphics[scale=0.8]{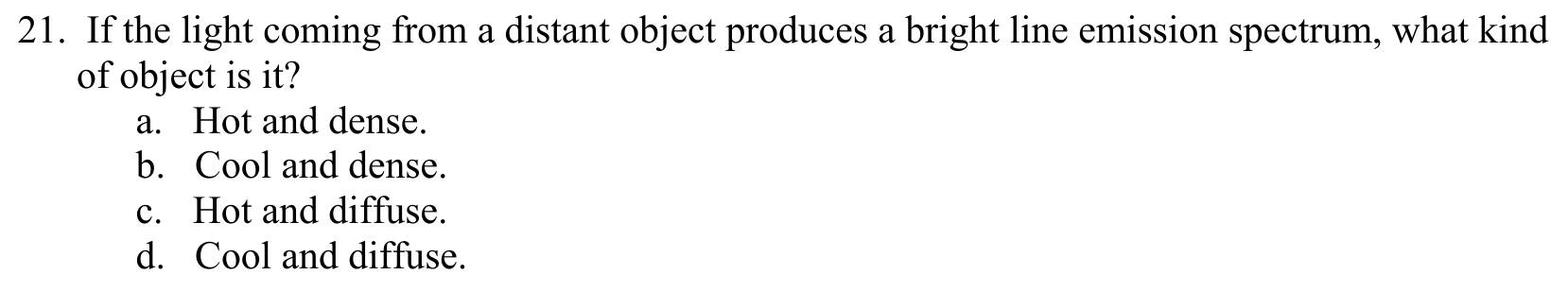}
\caption{\label{LSCI21}Item 21 from the LSCI.}
\end{figure*}

Before we proceeded with testing for potential violations of IRT's fundamental assumptions, we dropped two of the LSCI's items from our analysis: Items 21 and 25.
Item 21 (Fig.\ \ref{LSCI21}) had extremely poor goodness-of-fit statistics (e.g., $\chi^2 \approx 180$ with 22 degrees of freedom), regardless of the model used.  We found no clear relationship between student ability 
and success on Item 21.  We already suspected Item 21 might be inappropriate for this population based on our previous CTT analysis, which revealed that it had an extremely low 
discrimination value, which actually decreased from 0.14 to 0.12 pre- to post-instruction \cite{schlingman12}.  
Item 21 requires students to understand that a hot, diffuse cloud of gas produces a bright line emission spectrum and that a dense hot object does not, which distinguishes choices ``a" from ``c."
While 75\% of students post-instruction select either 
choice ``a" or 
choice ``c," over half of those students selected ``a," suggesting that many students do not understand the distinction between a ``dense" and a ``diffuse" object, even 
though they recognize that a ``bright line emission spectrum" must come from a hot object \cite{schlingman12}.  This item fails to probe the latent trait of interest since students'
responses are dominated by their knowledge of the definitions of these words. 

Item 25 had the largest difficulty parameter of any item on the LSCI ($b_{25} = 21$).  Item 25 presents students with graphs of energy output per second as a function of 
wavelength for four different objects (A-D); students must determine which object, if any, could be the same size as Object D.
The probability of a student correctly answering this item remains low across all abilities of students in the study population.  The reasoning required to correctly answer Item 25 challenges many professional astronomers, and we previously found its post-instruction CTT difficulty to 
be 0.89, with only 11\% of students giving the correct answer \cite{schlingman12}.  Because student success on this item was very weakly correlated with ability, it 
yielded essentially no useful information about students' abilities while degrading the overall goodness-of-fit of the data to the model.

After removing both Item 21 and Item 25, we examined whether or not we satisfied the two fundamental assumptions of IRT: local independence and unidimensionality.  If both of these assumptions hold, then the IRT model possesses the property of parameter invariance, which means that estimates of students' 
abilities do not depend on the specific items administered and estimates of item parameters do not depend on the abilities of students responding to those items
\cite{rupp06}.

\subsection{Local independence}
\label{LI}
An item is locally independent if the probability of correctly answering that item is entirely determined by a student's ability $\theta_p$ and not by his or her 
responses to other items or other sources of unaccounted-for variance \cite{embretson00}.  We used Yen's Q3 statistic to look for violations of local independence 
\cite{yen84}.  For each pair of items, Yen's Q3 statistic is the linear correlation between the items' residuals (i.e., the difference between students' observed and
3PL model-predicted scores).  If student ability $\theta_p$ is the only latent trait that determines the probability of correctly answering items, then there should
be essentially no correlation between the residuals of two different items.  Yen and Fitzpatrick recommend flagging item pairs for which the value of $|$Q3$| > 0.20$ 
\cite{yen06}.

\begin{figure*}
\includegraphics[scale=0.8]{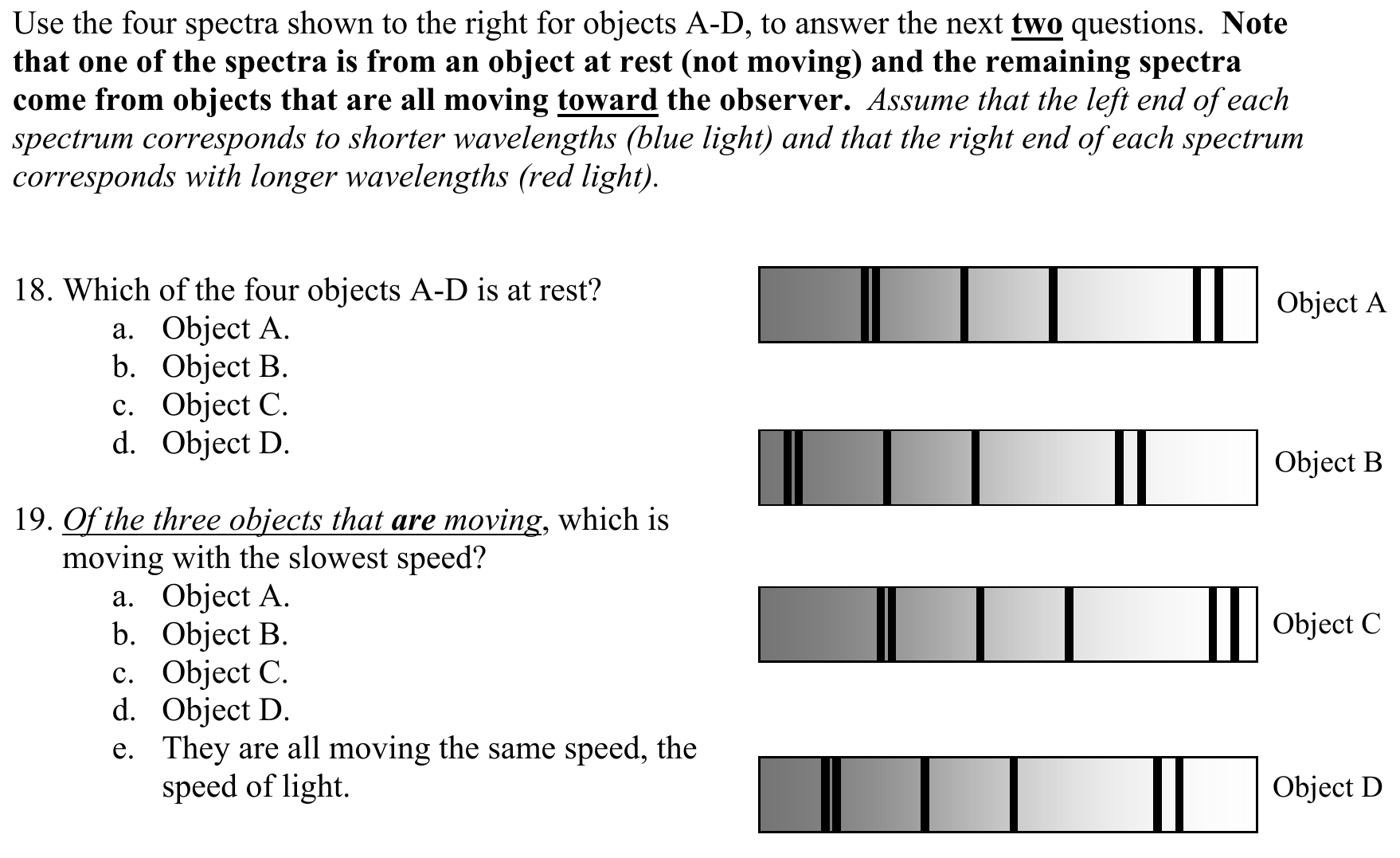}
\caption{\label{LSCI1819}Items 18 and 19 from the LSCI.}
\end{figure*}

We found that the following pairs of items had values of $|$Q3$| > 0.20$: Items 7 and 8, Items 18 and 19, and Items 2 and 22.  Before discussing how we dealt with
these violations of local independence, we must stress that just because these items have high Q3 values does not mean they are ``bad" items.  To the contrary,
Schlingman \emph{et al.}'s CTT analysis suggests that all of these items possess favorable psychometric properties \cite{schlingman12}.   If the flagged items
are not ``bad," then why do they have high Q3 values?   Take, for example, Items 18 and 19 (Fig.\ \ref{LSCI1819}).  Item 18 asks students to identify which of four spectra corresponds 
to an object at rest, while Item 19 asks students to identify which spectrum corresponds to the object moving the slowest toward the observer.  This item pair had a high 
Q3 value (Q3 = 0.51) because the probability of correctly answering Item 19 is not independent of the probability of correctly answering Item 18.  This pair of items exhibits what Yen calls ``item chaining," which means that one item builds off of the previous item such that knowing the answer to one item increases one's 
probability of correctly answering the other \cite{yen93}.  Someone who gives the correct answer to Item 18 has a much higher probability of giving the correct 
answer to Item 19, regardless of his or her ability level.   

The other item pairs with high Q3 values also exhibit item chaining.  Items 7 and 8 require students to determine which pictorial representation of the Bohr model of 
the atom corresponds to the formation of an absorption line and an emission line, respectively.  Items 2 and 22 ask students to reason about whether they can 
infer information about the color and temperature of a star, respectively, given its absorption line spectrum.  The high Q3 values for these pairs of items makes sense 
given the overlapping nature of their content.

We also found that Item 23 had high Q3 values with several items.  Item 23 asks students to compare the energy, frequency, wavelength, and speed of radio waves
and visible light.  The specific reasons why Item 23 exhibits local dependence with multiple items are not clear.  However, to correctly answer Item 23, students must 
synthesize their knowledge of how different types of light compare in terms of
energy, wavelength, frequency, and the speed at which they travel through a vacuum.  These ideas are so fundamental that students must frequently invoke them when reasoning about  other items on the LSCI. 

There are two possible solutions for how to deal with locally dependent item pairs.  One solution is to drop one item from each offending pair from the data set.
However, dropping items from the test reduces the amount of available information that can be used to estimate students' abilities.  We therefore took an alternative approach 
and combined each high-Q3 pair into a single polytomous item.  We tried several versions of the test with different pairwise combinations in an attempt to find a set of items that 
were all locally independent.  After several trials, we were able to resolve the problem for all items except for Item 23; regardless of the changes made to the rest of the test, this 
item was always found to be locally dependent on other items on the test.  We were therefore forced to remove Item 23 from the test.

We ended up with three polytomous items (Items 7 and 8 combined, Items 18 and 19 combined, and Items 2 and 22 combined).  These three items were calibrated using the two-parameter Graded Response Model \cite{samejima69}.  The Graded Response Model can be written as 
\begin{equation}
\label{grm}
P(X_{pi} \ge j|\theta_{p}, a_{i}, b_{ij}) = \frac{\exp [a_{i} (\theta_{p} - b_{ij})]}{1 + \exp [a_{i} (\theta_{p} - b_{ij})]} ~,
\end{equation}
where the student's ability $\theta_p$ and item's discrimination parameter $a_i$ have the same meaning as in Equations \ref{2pl} and \ref{3pl}.  Unlike the 2PL and 3PL models,
the Graded Response Model does not assign each item a single number $b_i$ to represent that item's difficulty.  Instead, each polytomous item is assigned multiple threshold 
parameters $b_{ij}$.  A given threshold parameter $b_{ij}$ represents the ability a student must have in order to have a 50\% probability of responding at or above the $j^{th}$ threshold
for a given item $i$.
For the three polytomous items we created out of Items 7 and 8, Items 18 and 19, and Items 2 and 22, there are two thresholds: $b_{i1}$ and $b_{i2}$, which represent the
ability a student needs in order to have a 50\% chance of scoring a 1 or 2, respectively.  See Embretson and Reise for a pedagogical treatment of the Graded Response
Model \cite{embretson00}.

Items 21, 23, and 25 were dropped from the instrument.  We maintained all other items in their original form and calibrated them using the 3PL model.  We will uses this twenty-item reduced version of the LSCI for all of the analyses subsequently described in this paper. Table \ref{q3} contains the matrix of Q3 values for every item pair on this reduced version
of the LSCI.  Table \ref{q3} shows that this version of the LSCI satisfies the assumption of local independence.

\begin{sidewaystable} \tiny
\centering
\caption{Yen's Q3 statistic for each pair of items.  Two items are considered locally independent if $|$Q3$| > 0.20$}.
\begin{tabular}{c c c c c c c c c c c c c c c c c c c c c}
\hline
\hline
 & Item 1 & Item 3  & Item 4 & Item 5 & Item 6 & Item 7 and 8  & Item 9 & Item 10  & Item 11 & Item 12  & Item 13 & Item 14 & Item 15 & Item 16 & Item 17 & Item 18 and 19 & Item 20 & Item 2 and 22 & Item 24 & Item 26 \\
\hline
Item 1	&	1.00	&	-0.04	&	-0.12	&	-0.07	&	-0.10	&	-0.07	&	-0.08	&	-0.03	&	-0.03	&	-0.15	&	-0.10	&	-0.08	&	0.19	&	-0.07	&	-0.04	&	-0.09	&	-0.05	&	-0.05	&	-0.09	&	-0.06 \\
Item 3	&		&	1.00	&	-0.02	&	-0.01	&	0.00	&	-0.02	&	0.02	&	0.07	&	-0.03	&	-0.02	&	0.00	&	-0.03	&	-0.01	&	-0.01	&	-0.02	&	0.01	&	0.04	&	-0.07	&	-0.02	&	0.03 \\
Item 4	&		&		&	1.00	&	-0.05	&	-0.05	&	-0.10	&	-0.09	&	-0.01	&	-0.03	&	-0.05	&	-0.03	&	-0.06	&	-0.09	&	-0.01	&	0.02	&	-0.03	&	-0.02	&	-0.03	&	-0.06	&	-0.05 \\
Item 5	&		&		&		&	1.00	&	-0.03	&	-0.04	&	-0.03	&	-0.04	&	-0.02	&	0.03	&	-0.02	&	0.07	&	-0.10	&	0.00	&	-0.04	&	-0.03	&	0.01	&	-0.09	&	-0.01	&	0.01 \\
Item 6	&		&		&		&		&	1.00	&	-0.08	&	-0.07	&	-0.05	&	-0.04	&	-0.04	&	-0.07	&	-0.01	&	-0.10	&	0.06	&	-0.03	&	-0.05	&	-0.01	&	-0.04	&	-0.06	&	0.04 \\
Item 7 and 8	&		&		&		&		&		&	1.00	&	-0.04	&	0.00	&	-0.02	&	-0.07	&	0.12	&	-0.09	&	-0.15	&	-0.08	&	-0.04	&	-0.05	&	-0.02	&	-0.03	&	-0.07	&	-0.01 \\
Item 9	&		&		&		&		&		&		&	1.00	&	-0.04	&	-0.06	&	0.16	&	-0.10	&	-0.04	&	-0.15	&	-0.09	&	-0.10	&	-0.05	&	-0.05	&	-0.12	&	0.14	&	-0.05 \\
Item 10	&		&		&		&		&		&		&		&	1.00	&	-0.05	&	-0.05	&	-0.02	&	0.00	&	-0.01	&	-0.05	&	-0.01	&	-0.02	&	0.08	&	-0.05	&	-0.05	&	-0.02 \\
Item 11	&		&		&		&		&		&		&		&		&	1.00	&	-0.06	&	-0.06	&	-0.05	&	-0.06	&	-0.05	&	-0.01	&	-0.01	&	-0.04	&	0.03	&	-0.04	&	-0.06 \\
Item 12	&		&		&		&		&		&		&		&		&		&	1.00	&	-0.07	&	0.02	&	-0.12	&	-0.04	&	-0.07	&	-0.03	&	-0.03	&	-0.18	&	0.16	&	-0.05 \\
Item 13	&		&		&		&		&		&		&		&		&		&		&	1.00	&	-0.01	&	-0.04	&	-0.02	&	0.03	&	-0.03	&	0.01	&	-0.12	&	-0.08	&	-0.03 \\
Item 14	&		&		&		&		&		&		&		&		&		&		&		&	1.00	&	0.05	&	0.00	&	0.00	&	-0.05	&	-0.03	&	-0.13	&	-0.05	&	-0.06 \\
Item 15	&		&		&		&		&		&		&		&		&		&		&		&		&	1.00	&	-0.01	&	-0.01	&	-0.07	&	0.01	&	-0.14	&	-0.18	&	-0.03 \\
Item 16	&		&		&		&		&		&		&		&		&		&		&		&		&		&	1.00	&	-0.02	&	0.01	&	-0.04	&	-0.08	&	-0.07	&	0.01 \\
Item 17	&		&		&		&		&		&		&		&		&		&		&		&		&		&		&	1.00	&	0.02	&	-0.01	&	0.09	&	-0.08	&	-0.07 \\
Item 18 and 19	&		&		&		&		&		&		&		&		&		&		&		&		&		&		&		&	1.00	&	0.02	&	-0.04	&	-0.05	&	-0.02 \\
Item 20	&		&		&		&		&		&		&		&		&		&		&		&		&		&		&		&		&	1.00	&	-0.10	&	-0.05	&	-0.03 \\
Item 2 and 22	&		&		&		&		&		&		&		&		&		&		&		&		&		&		&		&		&		&	1.00	&	-0.07	&	0.00 \\
Item 24	&		&		&		&		&		&		&		&		&		&		&		&		&		&		&		&		&		&		&	1.00	&	-0.03 \\
Item 26	&		&		&		&		&		&		&		&		&		&		&		&		&		&		&		&		&		&		&		&	1.00	\\																
\hline
\hline
\end{tabular}
\label{q3}
\end{sidewaystable}

\subsection{Unidimensionality}
\label{uni}
A test such as the LSCI is considered to be unidimensional if a single latent trait (aka ability $\theta_p$) can fully explain a student's performance on the test given the parameters describing the items
on that test (e.g., $a_i$, $b_i$, and $c_i$).  In other words, a test is unidimensional if it measures students' abilities on a single construct.  Local independence is a 
necessary but not sufficient condition for unidimensionality, so we conducted two additional tests to determine whether or not the assumption of unidimensionality 
holds.

For the first test, we fit the data with a two-latent-trait model and compared the results to those we obtained from the single-latent-trait model.  The two-dimensional
model did not yield a set of goodness-of-fit statistics that were better overall than those obtained by the unidimensional model.  Specifically, neither the average of the items' $\chi^2$ values nor the RMSEA were smaller for the two-dimensional model compared to the unidimensional model.  This suggests that a 
single latent trait is adequate to explain students' response patterns to the reduced version of the LSCI.

We then performed Bejar's test for unidimensionality \cite{bejar80}.  Bejar reasons as follows:  Imagine that a researcher suspects a test contains subsets of items 
that each probe their own unique construct.  The researcher could estimate item difficulties $b_i$ using the data for every item on the test.  The researcher could 
also estimate the item difficulties for the items on each subtest by using the data on those subtest items only.  If the test is truly unidimensional, then a plot of
the subtest-based item difficulty estimates versus the whole-test-based item difficulty estimates should show a series of points that fall near a line of slope one 
and intercept zero.  This is because the probability of correctly answering an item should not depend on which items are included on the test if the test is 
unidimensional.  Significant departures from this line are thus considered evidence that unidimensionality is violated.

For Bejar's test, we place items into three mutually-exclusive groups, which represented our hypothesis about which items might possibly form subtests that probe
different constructs.  One group included items that probe students' understandings of Wien's law and the luminosity-area-temperature relationship (Items 3, 6, 9, 12, 16, 20, 24, and 
26), another included items
that probe students' understandings of spectroscopy (Items 4, 7 and 8, 11, 17, 18 and 19, and 2 and 22), and the third included items that probed students' understandings of 
the properties of light (Items 1, 5, 10, 13, 14, and 15).  In all cases, the difficulty of each item fell within two standard errors of the target line of slope one and 
intercept zero.  We conclude that the results of Bejar's test are consistent with the assumption of unidimensionality.

\section{Results}
\label{results}

\subsection{Item parameters and model fit}
\label{items_fit}

Table \ref{dichotomous} shows the 3PL-estimated discriminations, difficulties, and guessing parameters of the dichotomous items on the reduced LSCI.  The standard errors
of these parameters are also shown.  In order to assess how well the 3PL model fit the data for each item, we grouped students into ability bins 0.1 logits wide, except in a few
cases where we had to increase the bin width in order to ensure there were at least five correct responses per bin.  A minimum of five correct responses per bin is generally considered sufficient to accurately estimate the average ability of a bin \cite{utts04}.  Some bins became extremely wide when we attempted to meet this criteria, so
we occasionally kept bin width at 0.1 logits and ignored all bins that failed to have at least five correct responses.  This is why some items have fewer degrees of 
freedom than others.
We compared each observed score 
to the expected score predicted by the 3PL model and calculated a $\chi^2$ statistic for each item.  Table \ref{dichotomous} also reports the $\chi^2$ values, the degrees of freedom, and the reduced $\chi^2$ values ($\chi_{r}^2$) for each item.

Table \ref{polytomous} contains the item parameters, their standard errors, the $\chi^2$ values, the degrees of freedom, and the reduced $\chi^2$ values for the three
polytomous items.  We calculated the $\chi^2$ values using the same procedure described above, except we found the expected score of each bin by taking 
the weighted average of the probability of receiving a score of 1 and a score of 2 (i.e., $P(X_{pi} = 1) + 2 P(X_{pi} = 2)$).

With a few exceptions, the $\chi_{r}^2$ values are close to unity, suggesting the IRT models adequately fit the data.  As an additional check on model fit, we plotted the model-predicted score on each item as a function of ability $\theta_p$; these plots are reproduced in
Appendix \ref{icc}.   In each plot, the black curve represents the model-predicted score while the red triangles represent the average scores of students in each bin.  
The overall close fit between the observed and predicted response patterns provides further
evidence that the IRT models are appropriate for modeling student ability.  

\begin{table*}[t] 
\centering
\caption{The discrimination ($a_i$), difficulty ($b_i$), and guessing parameters ($c_i$) of the seventeen dichotomous items from the reduced LSCI, along with their standard 
errors (SE).  The $\chi^2$, degrees of freedom (d.\ o.\ f.), and reduced $\chi^2$ ($\chi_{r}^2$) values are also shown.}
\begin{tabular}{c c c c c c c c c c c}
\hline
\hline
Item & $a_i$ & $a_i$'s SE & $b_i$ & $b_i$'s SE & $c_i$ & $c_i$'s SE & $\chi^2$ & d.\ o.\ f. & $\chi_{r}^2$  \\
\hline
Item 1	&	1.30	&	0.07	&	-0.75	&	0.05	&	0.00	&	0.00	&	31.66	&	15	&	2.11	\\
Item 3	&	1.71	&	0.33	&	2.09	&	0.14	&	0.20	&	0.01	&	16.40	&	18	&	0.91	\\
Item 4	&	1.81	&	0.22	&	-0.56	&	0.17	&	0.23	&	0.09	&	27.65	&	17	&	1.63	\\
Item 5	&	1.57	&	0.23	&	-0.28	&	0.21	&	0.44	&	0.07	&	11.40	&	18	&	0.63	\\
Item 6	&	1.83	&	0.21	&	0.19	&	0.10	&	0.24	&	0.04	&	20.05	&	18	&	1.11	\\
Item 9	&	2.22	&	0.24	&	0.70	&	0.05	&	0.16	&	0.02	&	23.42	&	17	&	1.38	\\
Item 10	&	1.33	&	0.21	&	1.22	&	0.10	&	0.23	&	0.03	&	21.61	&	19	&	1.14	\\
Item 11	&	1.81	&	0.22	&	0.89	&	0.07	&	0.21	&	0.02	&	26.95	&	19	&	1.42	\\
Item 12	&	2.68	&	0.36	&	0.18	&	0.08	&	0.36	&	0.03	&	18.99	&	17	&	1.12	\\
Item 13	&	1.35	&	0.17	&	0.32	&	0.13	&	0.18	&	0.05	&	22.29	&	19	&	1.17	\\
Item 14	&	1.34	&	0.23	&	-0.73	&	0.39	&	0.40	&	0.14	&	16.83	&	18	&	0.94	\\
Item 15	&	1.42	&	0.07	&	-0.32	&	0.04	&	0.00	&	0.00	&	37.48	&	14	&	2.68	\\
Item 16	&	1.36	&	0.24	&	0.35	&	0.18	&	0.36	&	0.06	&	11.44	&	18	&	0.64	\\
Item 17	&	1.26	&	0.22	&	1.88	&	0.13	&	0.15	&	0.02	&	25.94	&	18	&	1.44	\\
Item 20	&	1.14	&	0.23	&	0.99	&	0.15	&	0.31	&	0.05	&	17.76	&	19	&	0.93	\\
Item 24	&	2.57	&	0.33	&	0.53	&	0.06	&	0.31	&	0.02	&	15.87	&	18	&	0.88	\\
Item 26	&	2.33	&	0.34	&	1.39	&	0.07	&	0.23	&	0.01	&	30.07	&	19	&	1.58	\\
\hline
\hline
\end{tabular}
\label{dichotomous}
\end{table*}

\begin{table*}[t] 
\centering
\caption{The discrimination ($a_i$) and thresholds ($b_{i1}$ and $b_{i2}$) of the three polytomous items from the reduced LSCI, along with their standard 
errors (SE).  The $\chi^2$, degrees of freedom (d.\ o.\ f.), and reduced $\chi^2$ ($\chi_{r}^2$) values are also shown.}
\begin{tabular}{c c c c c c c c c c c}
\hline
\hline
Item & $a_i$ & $a_i$'s SE & $b_{i1}$ & $b_{i1}$'s SE & $b_{i2}$ & $b_{i2}$'s SE & $\chi^2$ & d.\ o.\ f. & $\chi_{r}^2$ \\
\hline
Item 7 and 8	&	1.06	&	0.05	&	-0.38	&	0.05	&	0.33	&	0.05	&	36.65	&	36	&	1.02	\\
Item 18 and 19	&	0.74	&	0.05	&	0.01	&	0.05	&	1.31	&	0.09	&	40.98	&	37	&	1.11	\\
Item 2 and 22	&	1.52	&	0.07	&	0.05	&	0.03	&	0.84	&	0.05	&	43.59	&	29	&	1.50	\\
\hline
\hline
\end{tabular}
\label{polytomous}
\end{table*}

\subsection{Item interpretations}
\label{items_interpret}

Before discussing the estimated student abilities and the learning gains achieved by different classes in our data set, we must comment on what the item parameters in
Tables \ref{dichotomous} and \ref{polytomous} tell us about the LSCI as an instrument.  First, consider the fact that the five items with the largest values of the discrimination
parameter $a_i$ (Items 12, 24, 26, 9, and 6, ranked from largest to smallest $a_i$) all come from the group of items that probe students' abilities to reason about and apply 
Wien's law and the luminosity-area-temperature relationship.  Furthermore, all five of these items require students to interpret a graph, such as star properties plotted on a graph of luminosity versus temperature.  The remaining items from the Wien/luminosity-area-temperature group (Items 3, 16, and 20) are entirely word-based questions and have lower discrimination 
values.  This demonstrates that graph-based items assessing Astro 101 students' understandings of Wien's law and/or the luminosity-area-temperature relationship are especially effective at 
discriminating between high- and low-ability students.

The plots in Appendix \ref{icc} show that a student must have an ability greater than 0 logits in order to have at least a 50\% probability of correctly answering any of the
Wien/luminosity-area-temperature items, with the exception of Items 6 and 12.  This is significant because the average post-instruction ability of students in the data set was set at
0 logits.  That means 50\% or more of students have less than a 50\% chance of giving the correct answer to six of the eight Wien/luminosity-area-temperature items even
at the end of their Astro 101 course.  

\begin{figure*}
\includegraphics[scale=0.8]{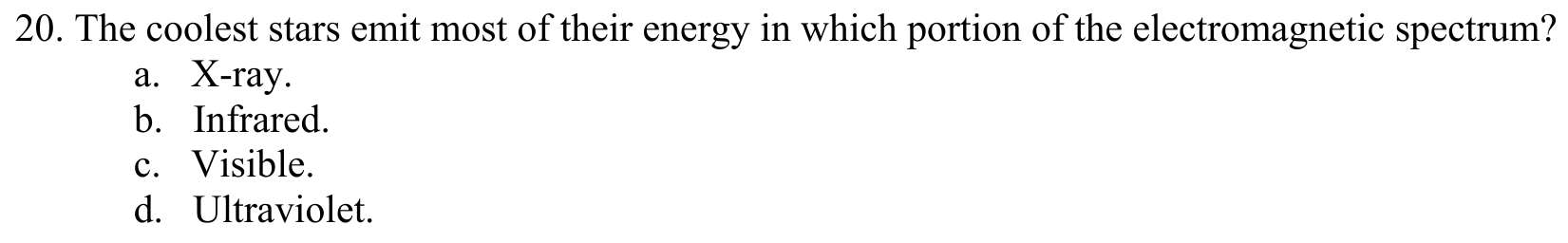}
\caption{\label{LSCI20}Item 20 from the LSCI.}
\end{figure*}

Overall, the Wien/luminosity-area-temperature items appear to be challenging for most Astro 101 students.  However, they are not so difficult that success on these items is unattainable,
which is why they tend to have high values of $a_i$, indicating that they are effective at discriminating between students of different abilities.  We suspect that many of these
items might have had higher discrimination values if not for the fact that they also have non-zero guessing parameters.  Items 3, 6, 9, and 26 have guessing parameters that
are around 0.20 to 0.25, which is consistent with low-ability students randomly guessing the correct answers when there are four to five available choices.  
Items 20 and 24 have guessing parameters $c_{20} = c_{24} = 0.31$ and Items 12 and 16 have $c_{12} = c_{16} = 0.36$.  These values of $c_i$ suggest that, after instruction, 
many low-ability students can eliminate at least one of the distractors before making a guess.  For example, only 10\% of students selected choice ``a" for Item 20 (Fig.\ \ref{LSCI20}), while 52\% selected ``b,"
24\% selected ``c," and 14\% selected ``d."  We conclude that the discriminatory powers of the Wien/luminosity-area-temperature items are attenuated because low-ability students have a non-zero probability of correctly guessing the correct answers.  This result is consistent with the findings of Wooten \emph{et al.}, which suggest that student performance on multiple-choice questions in many cases overestimates student understanding of a topic \cite{wooten14}.

In contrast to the Wien/luminosity-area-temperature items, items probing students' understandings of the properties of light (Items 1, 5, 10, 13, 14, and 15) tend to have both lower
discrimination values and lower difficulty parameters.  If we separate the nine dichotomous items with the largest values of $a_i$ from the nine dichotomous items with the 
smallest values of $a_i$, then we find that all of the properties of light items fall in the latter category.  Furthermore, many students with below average post-instruction abilities
($\theta_p < 0$ logits) still have a greater than 50\% chance of correctly answering Items 1, 5, 14, and 15.  Items 5 and 14 have extremely high guessing parameters (0.44 and 
0.40, respectively).  These questions ask students to select a photon (Item 5) or an electromagnetic wave (Item 14) with the largest energy.  We suspect that many students,
even those of low-ability, can eliminate one or more of the distractors based on what they learned in their Astro 101 classes about the relationships between the energy, 
wavelength, frequency, and color of light.  Items 1 and 15 are interesting because they both have guessing parameters of 0.  Both of these questions address the common
incorrect idea that more energetic forms of light travel faster.  The fact that these items have non-existent guessing parameters while simultaneous having low difficulties 
suggest that while many students can readily learn the fact that all forms of light travel at the same speed in a vacuum, low-ability students who never commit this fact to 
memory are almost certainly going to choose one of the distractors.  This implies that the distractors on these items are highly effective.  Overall, we are not surprised by the low difficulties and discriminatory capabilities of these items given
that they tend to probe what is simply declarative knowledge for many Astro 101 students.  

\subsection{Estimated student abilities}
\label{abilities}

\begin{figure*}
\includegraphics{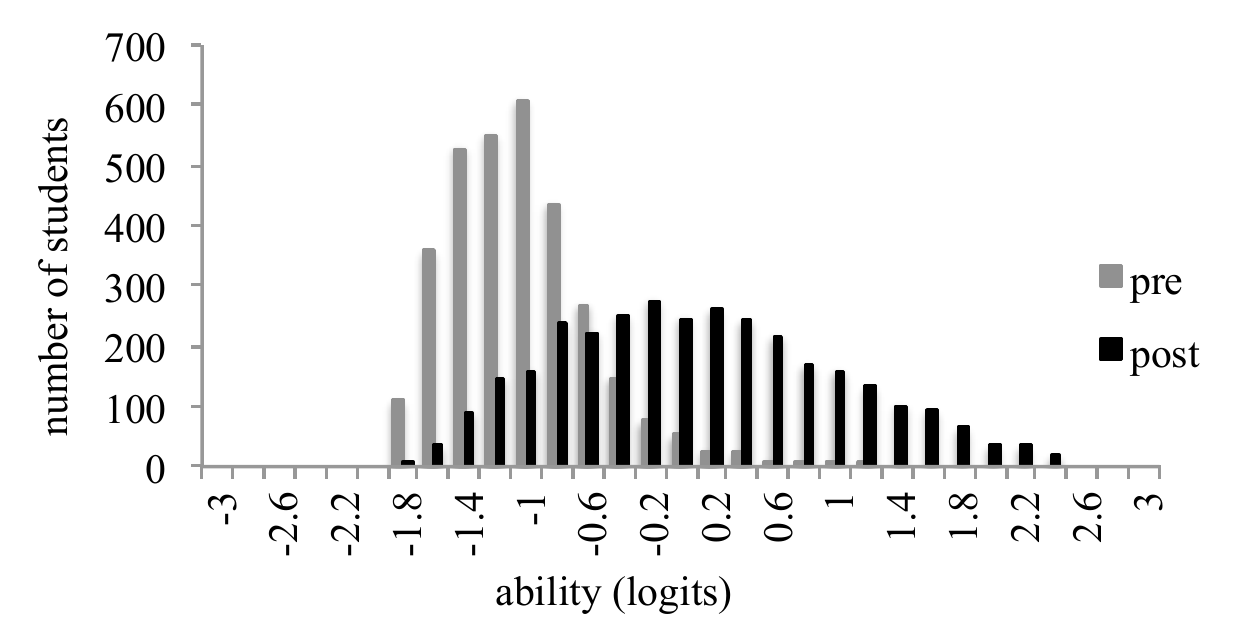}
\caption{\label{ability_dist}The distribution of pre- and post-instruction abilities for all 3205 students in the data set.}
\end{figure*}

Fig.\ \ref{ability_dist} shows the distribution of estimated pre- and post-instruction abilities for all 3205 students in the data set.  Pre-instruction abilities range from -1.9 to 1.0 logits, with an average of -1.1 logits and a standard deviation of 0.45 logits.  The post-instruction abilities span a wider range, from -1.9 to 2.4 logits, with an average of 0 logits and a standard deviation of 0.89 logits.  Recall that the average post-instruction ability is set at 0 logits, as described in Section \ref{data}.

\begin{figure*}
\includegraphics{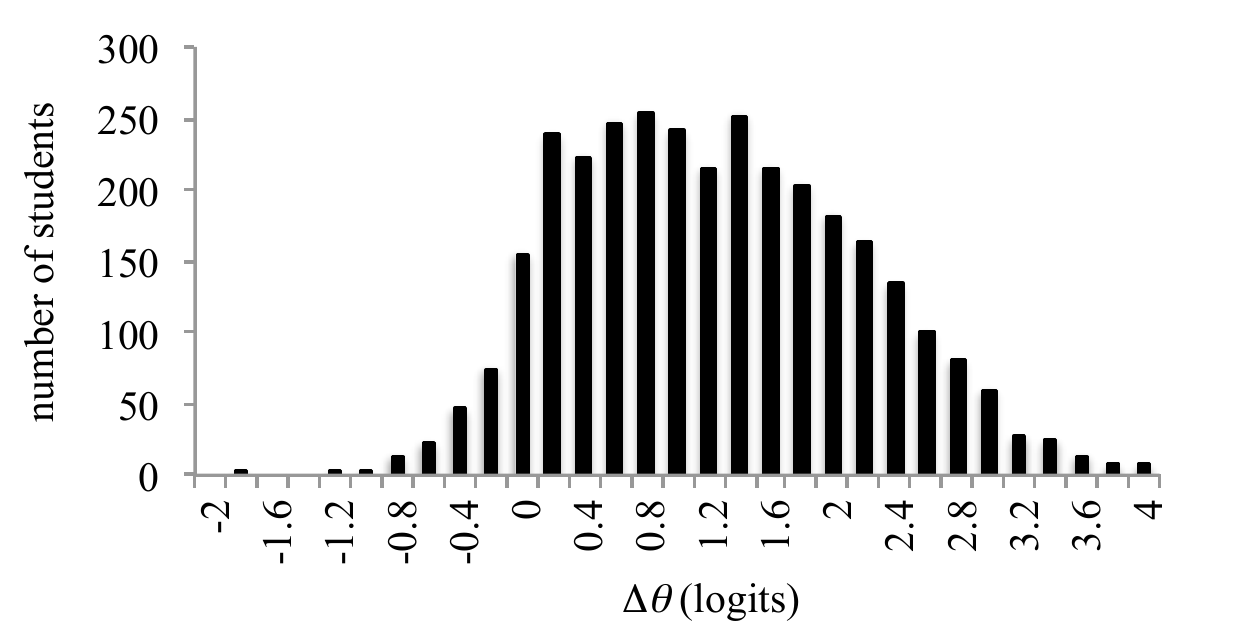}
\caption{\label{delta_ability_dist}The distribution of IRT-estimated gains ($\Delta \theta$) for all 3205 students in the data set.}
\end{figure*}

We calculated the difference between post- and pre-instruction abilities ($\Delta \theta$) for each student.  This difference represents an IRT-estimated learning gain \cite{wallace10}.  Fig. \ref{delta_ability_dist} shows the distribution of these IRT learning gains for all 3205 students.  The minimum ``gain" was -1.8 logits, the maximum was 4 logits, and the average was 1.1 logits with a standard deviation of 0.93 logits.  These data show a range in the shift of abilities, with less than 10\% of the assessed population exhibiting a shift $\leq 0$ logits, which would be consistent with students moving backward or achieving no improvement in their understanding.  

Since this study was carried out in order to examine the effects of active learning on individual students, we want to look at changes in abilities for students who took Astro 101 classes with high and low levels of interactive instruction.  The preceding study of Prather \emph{et al.} \cite{prather09} looked at the level of interactivity of classes in this data set with at least 25 students.  They estimated each class's level of interactivity based on instructors' responses to the Interactivity Assessment Instrument \cite{prather09}.  These responses allowed Prather \emph{et al.} to calculate an Interactivity Assessment Score (IAS) for each class.  IAS scores ranged from 0\% to 49\% and represent an estimate of the percentage of class time during which active learning techniques are used.  Prather \emph{et al} found that IAS scores of at least 25\% are necessary, but not sufficient, to produce classes with average normalized learning gains above $\langle g \rangle = 0.30$.  An average normalized gain of $\langle g \rangle = 0.30$ is significant because Hake \cite{hake98} found that only interactive physics classes -- and not traditionally taught classes -- were able to achieve this level of improvement in student performance on the Force Concept Inventory (FCI).  Consequently, Hake defined $\langle g \rangle = 0.30$ as the cutoff between ``low" and ``medium" levels of gain. 

\begin{figure*}
\includegraphics{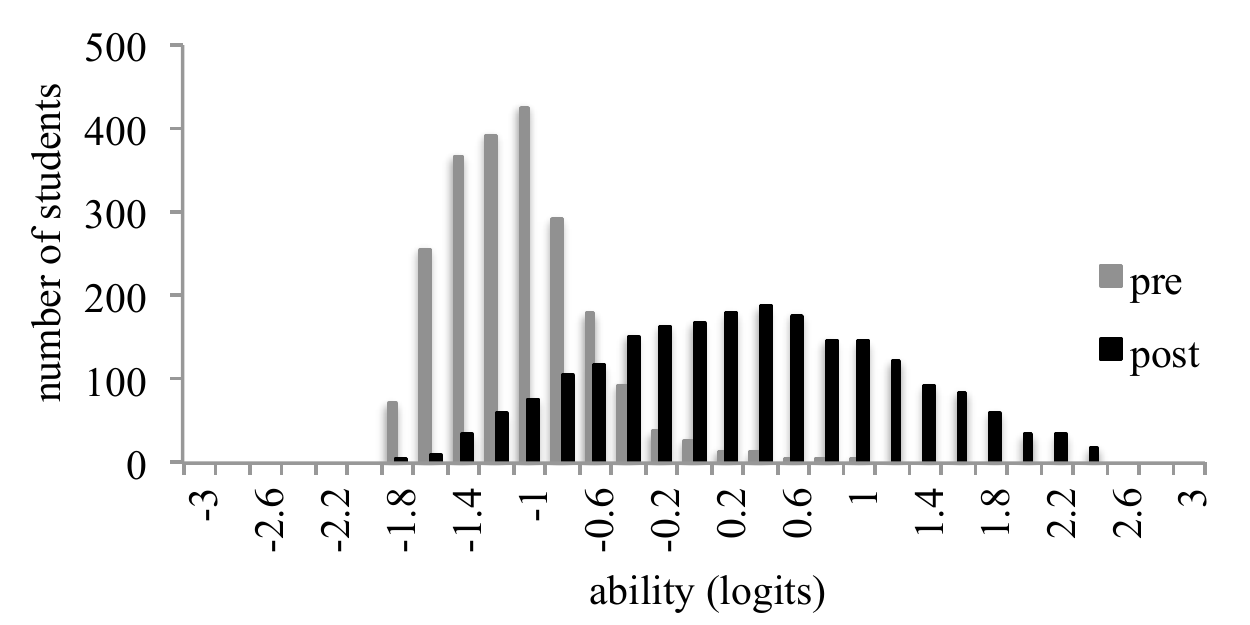}
\caption{\label{highIAS_dist}The distribution of pre- and post-instruction abilities for the 2178 students in high-IAS classes with at least 25 students.}
\end{figure*}

\begin{figure*}
\includegraphics{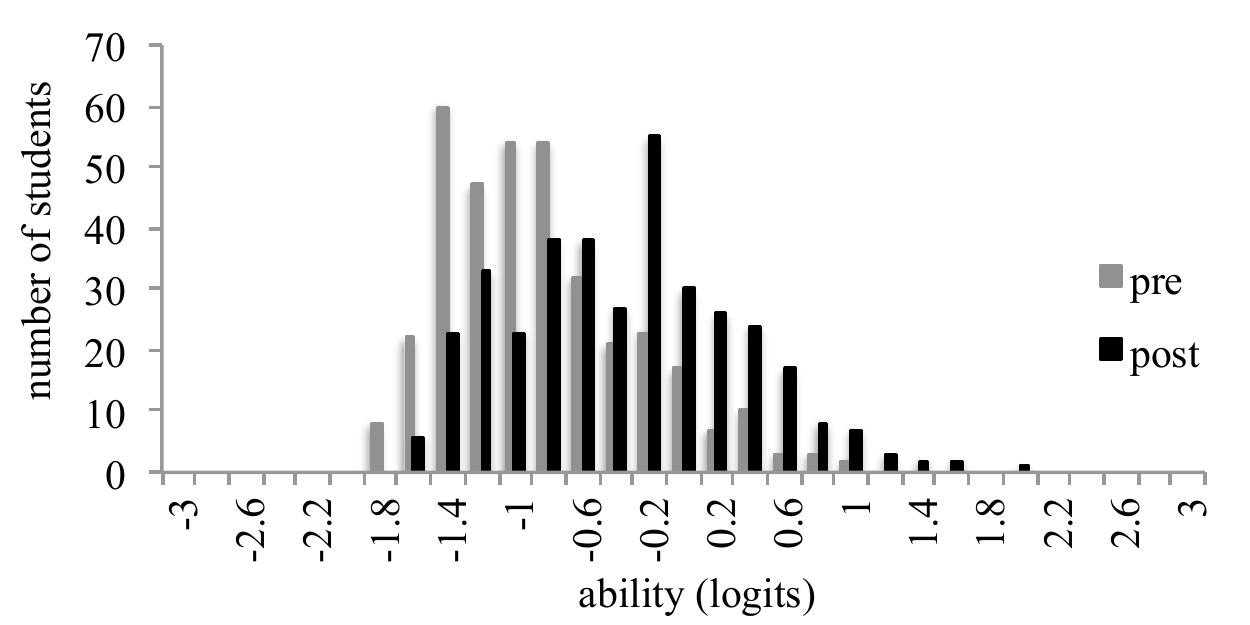}
\caption{\label{lowIAS_dist}The distribution of pre- and post-instruction abilities for the 363 students in low-IAS classes with at least 25 students.}
\end{figure*}

In this study, we again look at at students enrolled in Astro 101 classes with at least 25 students.  Like the prior study by Prather \emph{et al.}, we divide these classes into two groups: high-IAS (i.e., IAS $\ge$ 25\%) and low-IAS (IAS $<$ 25\%).  Figs. \ref{highIAS_dist} and \ref{lowIAS_dist} show the pre- and post-instruction ability distributions for students in high- and low-IAS classes, respectively.  Students in high-IAS classes have pre-instruction abilities that range from -1.9 logits to 0.92 logits with an average of -1.2 logits and a standard deviation of 0.42 logits.  Their post-instruction abilities range from -1.9 logits to 2.4 logits with an average of 0.23 logits and a standard deviation of 0.88 logits.  In contrast, students in low-IAS classes have pre-instruction abilities that range from -1.9 logits to 0.91 logits with an average of -0.95 logits and a standard deviation of 0.56 logits.  Their post-instruction abilities range from -1.8 logits to 1.9 logits with an average of -0.45 logits and a standard deviation of 0.69 logits.  While the distributions of pre- and post-instruction abilities for students in high- and low-IAS classes cover approximately the same range, the post-instruction averages are noticeably different.  A one-tailed t-test for two independent samples revealed that the difference in the post-instruction means was statistically significant ($p < 0.0001$) and of large effect size (Cohen's $d = 0.80$) \cite{cohen88}.  Interestingly -- and unexpectedly -- a one-tailed t-test also revealed that the difference in the pre-instruction means was also statistically significant ($p < 0.0001$) and of medium effect size (Cohen's $d = 0.49$).  These results show that even though the population of students from high-IAS classes began with a smaller average pre-instruction ability, they had a higher average post-instruction ability than their peers in low-IAS classes.  A significant number of students in high-IAS classes moved into regions of ability space that were unoccupied in the pre-instruction distribution, and they did so at a greater percentage than students in low-IAS classes.  
This means a significant number of students in high-IAS classes, compared to students in low-IAS classes, acquired astronomical reasoning abilities and knowledge that were not held by most students prior to instruction.
This important result is consistent with the CTT findings of Prather \emph{et al.} and with decades of research highlighting the pedagogical effectiveness of interactive instruction \cite{freeman14}.

\begin{figure*}
\includegraphics{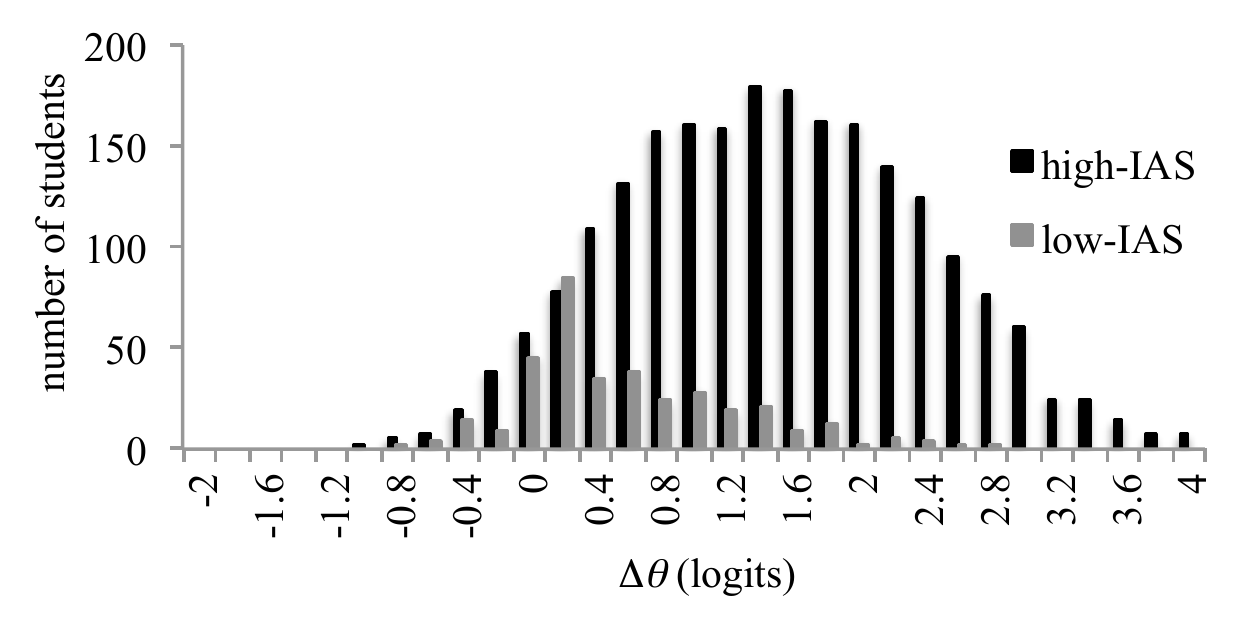}
\caption{\label{IAS_delta_dist}The distribution of IRT-estimated gains ($\Delta \theta$) for the 2178 students in high-IAS classes and the 363 students in low-IAS classes.}
\end{figure*}

The disparity in student achievement between high- and low-IAS classes is also seen when we examine the distributions of IRT-estimated learning gains ($\Delta \theta$) (see Fig.\ \ref{IAS_delta_dist}).  Students in high-IAS classes have values of $\Delta \theta$ that range from -1.2 logits to 4 logits with an average of 1.4 logits and a standard deviation of 0.90 logits.  Students in low-IAS classes have values of $\Delta \theta$ that range from -0.95 logits to 2.6 logits with an average of 0.49 logits and a standard deviation of 0.66 logits.  Once again, a one-tailed t-test revealed the difference in these averages to be statistically significant ($p < 0.0001$).  
This difference in averages also corresponds to a very large effect size (Cohen's $d = 1.2$), according to the effect size classification scheme proposed by Sawilowsky 
\cite{sawilowsky09}. 
Surprisingly, students in high-IAS classes averaged a pre-post improvement in their abilities that was almost an entire logit greater than the average pre-post ability improvement of students in low-IAS classes.  To get a sense of the meaning of a difference of 1 logit, consider the difficulty parameters of the seventeen dichotomous items on the LSCI (Table \ref{dichotomous}).  These difficulty parameters range from -0.75 logits to 2.09 logits.  A student whose ability increases by 1 logit will have a significantly higher probability of correctly answering many of the LSCI's items.  This same reasoning also applies to the three polytomous items (Table \ref{polytomous}).  For example, a student with an ability of 0 logits has a 55\% chance of answering item 6 correctly, but a student with an ability 1 logit greater has an 85\% probability of answering this item correctly.  

\begin{figure}
\includegraphics{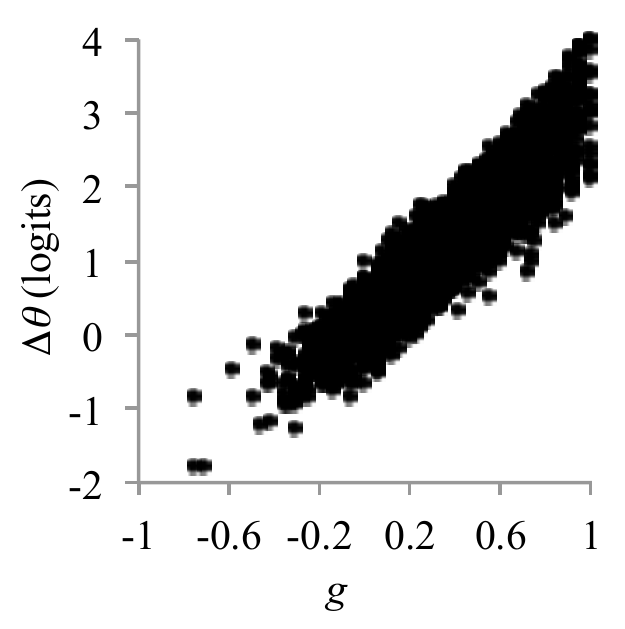}
\caption{\label{thetag} $\Delta \theta$ versus $g$ for all 3205 students in the LSCI national data set.}
\end{figure}

Many astronomy and physics education researchers frequently use Hake's average normalized gain $\langle g \rangle$ to make inferences about the amount of learning experienced by populations of students \cite{lindell01, williamson13, finkelstein05, hake98}, including the earlier study by Prather \emph{et al.} \cite{prather09}.  In addition to calculating $\Delta \theta$ for all 3205 students in the data set, we also calculated their normalized gains $g$.  Fig.\ \ref{thetag} shows a graph of $\Delta \theta$ versus $g$ for all 3205 students.  There is a definite correlation between the two measures ($r =$ 0.93).  But note that each value of $g$ corresponds to a range of values of $\Delta \theta$ approximately 1 logit or more wide.  There are many IRT-estimated gains associated with a single value of $g$.  Two students who have values of $\Delta \theta$ separated by 1 logit have experienced significantly different improvements in their underlying abilities, even if they possess the same normalized gain.  This result makes sense when one recalls that IRT estimates a person's ability based on the relative difficulty of the questions she correctly answered, not just the total number of correct answers.  This result suggests that while average normalized gains may be good at summarizing the performance of a population of students, $g$ may not be as informative an indicator of the learning gains of individual students. 

Of course, Fig.\ \ref{thetag} also shows that there are also multiple values of $\Delta \theta$ for each value of $g$.  Why then do we claim that $\Delta \theta$ more robustly models changes in student understanding than $g$?  Recall that in order to obtain these values of $\Delta \theta$ we had to perform numerous statistical tests to demonstrate that the IRT models we used fit the data and satisfied the underlying assumptions of local independence and unidimensionality.  When these conditions are satisfied, IRT models 
provide estimates of students' abilities that are independent of the specific items they answered \cite{rupp06}.  If one wants to argue that $g$ is a more accurate measure than $\Delta \theta$ of learning gains, then one must justify why a raw test score is a better measure than $\theta$ of the latent trait of student ability, despite of the amount of statistical rigor required to produce $\theta$ values.  We believe that the statistical analysis underlying $\Delta \theta$ makes it highly unlikely that $g$ is a superior measure of learning gain.  

\subsection{Effectiveness of Different Astro 101 Classes}
\label{gains}

Earlier investigations of the LSCI national data set examined the relationships between the average normalized learning gains of classes, the amount of time devoted to active learning, and the quality of an instructor's implementation of those strategies \cite{prather09, rudolph10}.  Consequently, we are interested in examining the 
average IRT-estimated learning gains for classes in this data set, as well as the pre- and post-instruction ability histograms for individual classes, in order to determine how 
well each class did with regards to evolving students' underlying abilities over the course of the semester.

\begin{figure}
\includegraphics{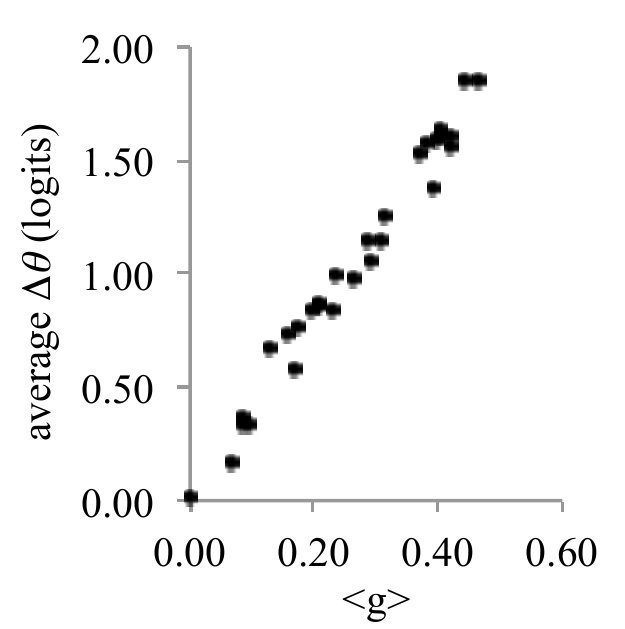}
\caption{\label{theta_g} Average $\Delta \theta$ versus $\langle g \rangle$ for classes with at least 25 students in the LSCI national data set.}
\end{figure}

\begin{figure}
\includegraphics{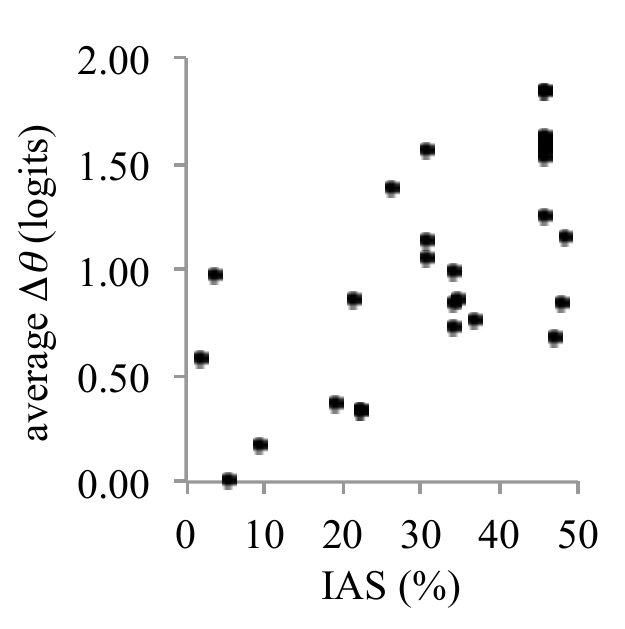}
\caption{\label{theta_IAS} Average $\Delta \theta$ versus IAS for classes with at least 25 students in the LSCI national data set.}
\end{figure}

For each class with at least 25 students in the data set, Table \ref{class_info} shows the type of institution at which the class was taught, the number of enrolled students, the average 
pre- and post-instruction scores on the LSCI, $\langle g \rangle$, the average pre- and post-instruction abilities, average $\Delta \theta$, and the instructor's IAS.  The classes are ordered from largest to smallest average $\Delta \theta$.  Fig.\ \ref{theta_g} plots these average $\Delta \theta$ values versus $\langle g \rangle$.  There is a large correlation between these two measures ($r = 0.99$), which supports the robustness of the results reported in Prather \emph{et al.} \cite{prather09}, Rudolph \emph{et al.} \cite{rudolph10}, and Schlingman \emph{et al.} \cite{schlingman12}.  Fig.\ \ref{theta_IAS} shows average $\Delta \theta$ versus IAS.  As expected, this reveals that spending more time on active learning strategies is necessary to move beyond
small learning gains, supporting the validity of the findings of prior CTT studies, which often report measures of learning gain that are two times larger for students taught interactively than students taught traditionally \cite{prather09, hake98}.  A similar plot of $\langle g \rangle$ versus IAS in Prather \emph{et al.} \cite{prather09} revealed that only instructors with an IAS greater than or equal to 25\% had classes with at least a medium gain ($\langle g \rangle = 0.3$ according to Hake \cite{hake98}).  Comparing Fig.\ \ref{theta_IAS} with the $\langle g \rangle$ versus IAS plot from Prather \emph{et al.} \cite{prather09} suggests that an average $\Delta \theta = 1$ is approximately equivalent to $\langle g \rangle = 0.3$.  Fig.\ \ref{theta_IAS} also suggests that an IAS of 25\% is a necessary, though not sufficient, condition to achieve average $\Delta \theta > 1$.  This result strongly suggests that simply making a class more interactive is not enough to maximize student learning; the quality of an instructor's ability to create an effective active learning classroom plays a significant role in student learning outcomes.  

\begin{table*} \scriptsize
\centering
\caption{The institution type, number of enrolled students, average pre- and post-instruction LSCI scores, average normalized gain $\langle g \rangle$, average pre- and post-instruction
abilities $\theta$, average IRT learning gain $\Delta \theta$, and IAS for each class in the national data set with at least 25 students. }
\begin{tabular}{c c c c c c c c c c}
\hline
\hline
Class & Institution Type & Students & Average Pre-Score & Average Post-Score & $\langle g \rangle$ & Average Pre-$\theta$ & Average Post-$\theta$ & Average $\Delta \theta$ & IAS  \\
\hline
Class 1	&	Research University	&	96	&	5.79	&	15.26	&	0.47	&	-1.18	&	0.66	&	1.84	&	45.9	\\
Class 2	&	Research University	&	93	&	6.06	&	14.92	&	0.44	&	-1.16	&	0.68	&	1.84	&	45.9	\\
Class 3	&	Research University	&	63	&	6.35	&	14.68	&	0.42	&	-1.04	&	0.57	&	1.61	&	45.9	\\
Class 4	&	4-yr Masters/Bach.\ Univ.	&	33	&	6.94	&	15.00	&	0.42	&	-0.96	&	0.60	&	1.56	&	30.8	\\
Class 5	&	Research University	&	84	&	5.85	&	13.99	&	0.40	&	-1.15	&	0.48	&	1.63	&	45.9	\\
Class 6	&	Research University	&	65	&	5.17	&	13.58	&	0.40	&	-1.19	&	0.41	&	1.60	&	45.9	\\
Class 7	&	Research University	&	444	&	5.41	&	13.61	&	0.40	&	-1.21	&	0.37	&	1.58	&	45.9	\\
Class 8	&	4-yr Masters/Bach.\ Univ.	&	43	&	6.79	&	14.37	&	0.39	&	-0.98	&	0.40	&	1.38	&	26.1	\\
Class 9	&	Research University	&	344	&	5.21	&	13.20	&	0.38	&	-1.24	&	0.33	&	1.57	&	45.9	\\
Class 10	&	Research University	&	61	&	5.33	&	13.02	&	0.37	&	-1.26	&	0.27	&	1.53	&	45.9	\\
Class 11	&	Research University	&	402	&	5.94	&	12.27	&	0.32	&	-1.12	&	0.14	&	1.25	&	45.9	\\
Class 12	&	4-yr Masters/Bach.\ Univ.	&	36	&	6.53	&	12.50	&	0.31	&	-1.01	&	0.13	&	1.14	&	30.8	\\
Class 13	&	4-yr Masters/Bach.\ Univ.	&	28	&	7.89	&	13.14	&	0.29	&	-0.83	&	0.22	&	1.05	&	30.8	\\
Class 14	&	2-yr College	&	66	&	5.58	&	11.41	&	0.29	&	-1.16	&	-0.02	&	1.14	&	48.6	\\
Class 15	&	Research University	&	64	&	6.16	&	11.45	&	0.27	&	-1.04	&	-0.07	&	0.97	&	3.6	\\
Class 16	&	4-yr Masters/Bach.\ Univ.	&	40	&	5.65	&	10.43	&	0.23	&	-1.19	&	-0.19	&	0.99	&	34.3	\\
Class 17	&	4-yr Masters/Bach.\ Univ.	&	40	&	6.55	&	11.05	&	0.23	&	-1.06	&	-0.22	&	0.84	&	47.8	\\
Class 18	&	4-yr Masters/Bach.\ Univ.	&	65	&	6.23	&	10.38	&	0.21	&	-1.08	&	-0.22	&	0.86	&	21.6	\\
Class 19	&	Research University	&	47	&	6.62	&	10.62	&	0.21	&	-1.04	&	-0.18	&	0.86	&	34.5	\\
Class 20	&	4-yr Masters/Bach.\ Univ.	&	41	&	5.49	&	9.59	&	0.20	&	-1.25	&	-0.41	&	0.84	&	34.3	\\
Class 21	&	4-yr Bachelors College	&	33	&	5.24	&	8.91	&	0.18	&	-1.24	&	-0.48	&	0.76	&	36.9	\\
Class 22	&	Research University	&	28	&	5.93	&	9.36	&	0.17	&	-0.94	&	-0.36	&	0.58	&	2.1	\\
Class 23	&	4-yr Masters/Bach.\ Univ.	&	42	&	5.17	&	8.45	&	0.16	&	-1.27	&	-0.54	&	0.73	&	34.3	\\
Class 24	&	4-yr Masters/Bach.\ Univ.	&	77	&	5.21	&	7.88	&	0.13	&	-1.26	&	-0.59	&	0.67	&	47.3	\\
Class 25	&	2-yr College	&	27	&	5.37	&	7.41	&	0.10	&	-1.16	&	-0.83	&	0.33	&	22.1	\\
Class 26	&	4-yr Masters/Bach.\ Univ.	&	62	&	5.85	&	7.61	&	0.09	&	-1.13	&	-0.76	&	0.37	&	19.3	\\
Class 27	&	2-yr College	&	25	&	6.00	&	7.72	&	0.09	&	-1.19	&	-0.86	&	0.33	&	22.1	\\
Class 28	&	4-yr Bachelors College	&	27	&	5.26	&	6.74	&	0.07	&	-1.21	&	-1.05	&	0.16	&	9.7	\\
Class 29	&	4-yr Masters/Bach.\ Univ.	&	65	&	10.12	&	10.14	&	0.00	&	-0.28	&	-0.27	&	0.01	&	5.4	\\
\hline
\hline
\end{tabular}
\label{class_info}
\end{table*}

We created pre- and post-instruction ability histograms
for each of the twenty-nine classes with at least 25 students.  These histograms are located in Appendix \ref{histograms}.  We now move to investigate these histograms of student abilities in order to gain deeper insights into the effectiveness of instruction in different classes.  We will focus our investigation on the distributions from three classes that represent dramatically different outcomes.

\begin{figure}
\includegraphics{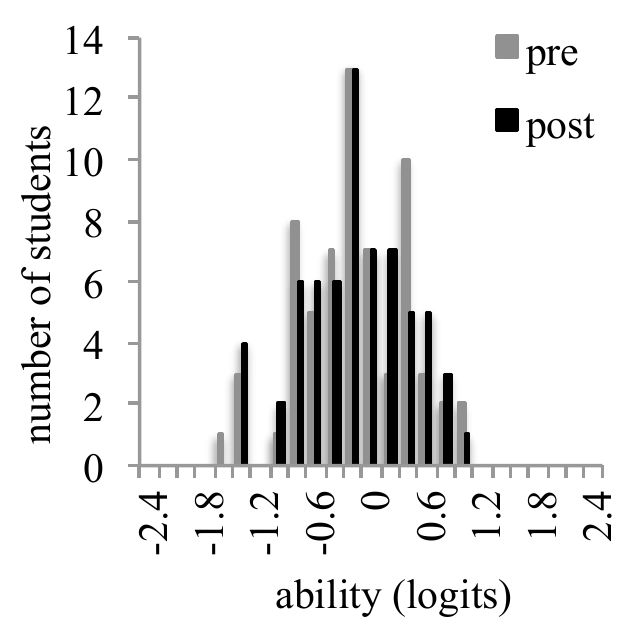}
\caption{\label{class29_main}The pre- and post-ability histogram for Class 29.}
\end{figure}

Class 29 (Fig.\ \ref{class29_main}) has the lowest average $\Delta \theta$ value (0.01) and the third lowest IAS in the data set.  Note that Class 29 also has the highest average pre-instruction score on the 
LSCI and the highest average pre-instruction ability.  Despite the apparent advantages this class of students had at the beginning of their Astro 101 course, the histogram graphically illustrates that very few students improved in ability since the pre- and post-instruction distributions almost completely overlap one another.

\begin{figure}
\includegraphics{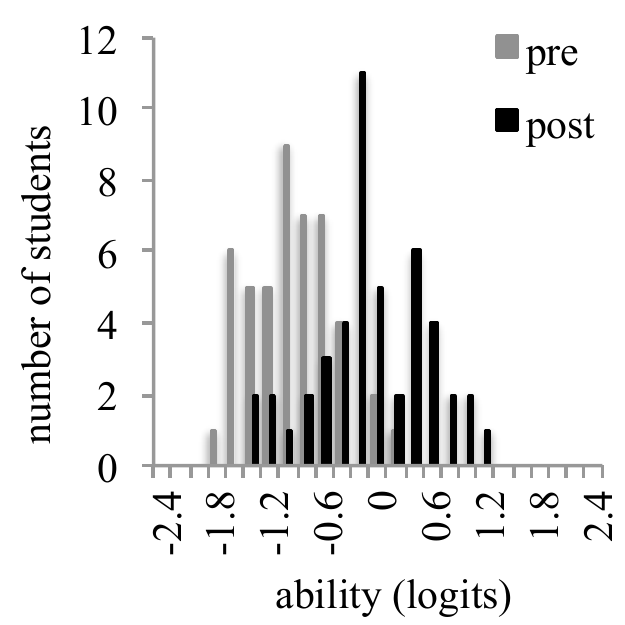}
\caption{\label{class19_main}The pre- and post-ability histograms for Class 19.}
\end{figure}

In contrast, the histogram for Class 19 (Fig.\ \ref{class19_main}) shows that the pre-instruction and post-instruction distributions of student abilities have far less overlap than what we observe for Class 29.  It is important to note that there are a considerable number of students with post-instruction abilities that none of the students had prior to instruction.   This is powerful and illustrative evidence for the assertion that significant learning did occur in Class 19.  During society meetings and colloquia talks, and as part of our professional development workshops, we have asked faculty to compare the pre- and post-instruction ability distributions for Classes 19 and 29.  It is common for faculty to verbally express how impressed they are with the learning in Class 19 -- often stating that they would be pleased if their own classes achieved a similar shift from pre- to post-instruction abilities.  However, Class 19 does not represent the upper limit of what we observed with respect to student learning.  While there is a clear separation between the distributions of the pre- and post-instruction abilities, there is also a significant amount of overlap in these ability values.  This suggests that there may be some students who did not experience any improvement in their abilities as a result of the instruction from Class 19.  Furthermore, the majority of post-instruction abilities have values less than 0 logits.  This means that many students in Class 19 still had post-instruction abilities that were below the study post-instruction average.   As faculty become aware of the relatively low post instruction abilities of the students in Class 19, we have found it valuable to show them data from a class with very little overlap in the pre-post ability distributions, an impressive average $\Delta \theta$, and which has students who have achieved high post-instruction abilities.
 
\begin{figure}
\includegraphics{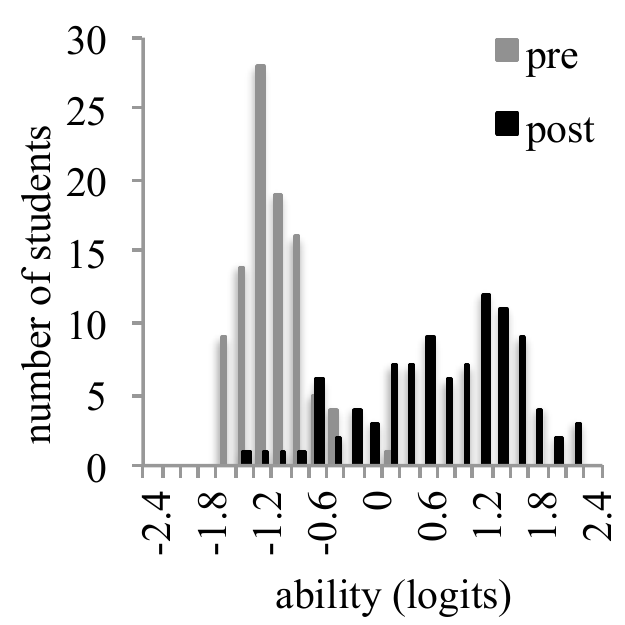}
\caption{\label{class1_main}The pre- and post-ability histograms for Class 1.}
\end{figure}

The most impressive shift in student abilities was observed with Class 1 (Fig.\ \ref{class1_main}). There is astonishingly little overlap between the distributions of students' pre- and post-instruction abilities for these classes.  A careful inspection of the pre- and post-instruction distributions for Class 1 also reveals that after instruction almost every student has an ability that none of the students had prior to instruction -- a truly remarkable teaching and learning accomplishment.  Additionally, most students in Class 1 have post-instruction abilities greater than 0 logits, meaning they were above the data set's post-instruction average.  Some students in Class 1 achieved post-instruction abilities of 2.2 logits, which is at the extreme high end of the distributions shown in Fig.\ \ref{ability_dist} -- and this in a class with one of the lower average pre-instruction abilities.  Overall, Class 1 serves as an example for how transformative a single semester introductory astronomy course can be with regards to improving students' conceptual and reasoning abilities on fundamental astrophysical ideas.  During our presentations, after sharing the results from Class 1 with faculty, most are quick to switch to aspiring for learning outcomes similar to Class 1 over Class 19.       

Even though there is a large correlation between average $\Delta \theta$ values and $\langle g \rangle$, Fig.\ \ref{theta_g} and Table \ref{class_info} also show that it is possible for
two classes to have the same value for $\langle g \rangle$ but very different average $\Delta \theta$ values, and vice versa.  For example, Classes 3 and 8 have similar values of $\langle g \rangle$ (0.42 and 0.39, respectively) but $\Delta \theta$ values that differ by 0.23 logits (1.61 logits versus 1.38 logits, respectively). We also suspect it is possible to have a class with a large average $\Delta \theta$ value but a histogram of pre- and post-instruction ability distributions that is unimpressive in important aspects (e.g., most students are still below average post-instruction average $\theta$).  Such findings as these reinforce the value of an IRT analysis for extracting information from larger educational data sets.  The above considerations, plus our above analyses of Class 1, 19, and 29, suggest that, instructors seeking a full understanding of the effectiveness of their classroom instruction should compare a measure of their classes' average improvement (e.g., $\langle g \rangle$ and/or average $\Delta \theta$) with the distribution of students' pre- and post-instruction abilities, and the distribution of individual student gains $\Delta \theta$.  By combining these multiple perspectives on individual and classwide abilities and gains one can obtain a much more robust understanding of the effects of instruction.  Even so, the outcomes of one class are much more meaningful when compared to the outcomes of other classes; using a widely validated and applied instrument such as the LSCI allows instructors to understand the efficacy of their teaching in both local and global contexts.

\section{Summary and Conclusions}
\label{conclusions}

We used IRT to analyze the responses of 3205 Astro 101 students from sixty-nine classes (representing all types of colleges and universities) to the LSCI.  As part of
our analysis, we removed two items from the LSCI: Item 21, due to the fact that it is known to be a problematic item \cite{schlingman12}, and Item 25, since it is so difficult that
students' success on it shows only a weak correlation with their underlying abilities.  In order to satisfy IRT's assumption of local independence, we removed a third item, Item 
23, and we combined three pairs of items (Items 7 and 8, Items 18 and 19, and Items 2 and 22) into three polytomous items.  After making these modifications, we were able to
fit the 3PL model to the remaining seventeen dichotomous items and the Graded Response Model to the three polytomous items, while simultaneously satisfying IRT's 
assumptions of local independence and unidimensionality.  By satisfying these assumptions -- and in contrast to classical test theory (CTT) -- we achieved parameter invariance, which means our estimates of students' 
underlying abilities and the parameters of the items to which they responded do not depend upon one another \cite{rupp06}. 

Our IRT analysis provided new insights into the functioning of many of the LSCI's items.  Since the 3PL model contains a ``guessing parameter" ($c_i$), the probability of correctly answering an item with a large value of $c_i$ (e.\ g.\ Item 3) is influenced by many low-ability students guessing the correct answer.  Items with small values of $c_i$ (e.\ g.\ Item 1) must possess particularly powerful distractors that limit the influence of guessing on the probability of students getting the right answer.  This kind of analysis 
is not possible with CTT.  

When we look at specific categories of items on the LSCI, we learned that items probing the properties of light are the easiest
for students to correctly answer.  In contrast, items that require students to reason using Wien's law and/or the luminosity-area-temperature relationship are among the most difficult and 
discriminating items of the LSCI, especially when these items require students to interpret graphical or pictorial representations.

The results of our IRT analysis also support the robustness of the research results from prior classical test theory (CTT) analyses of this data set \cite{prather09, rudolph10,schlingman12}.  We split 
all classes with at least 25 students in the data set into two categories: classes in which the instructor used active learning strategies for 25\% of class time or more and classes
in which the instructor spent less than 25\% of class time using active learning strategies.  Students in classes that used active learning strategies for 25\% of class time or more had 
higher average post-instruction abilities and larger average IRT-estimated learning gains (average $\Delta \theta$) than students in classes that spent less time on active learning 
strategies -- despite the fact that the higher IAS classes actually began Astro 101 with lower pre-instruction abilities.  Students in high IAS classes had an average $\Delta \theta$ 
that was approximately 1 logit greater than their peers in low IAS classes.  This difference of 1 logit represents a significant fraction of the range of the LSCI's items' difficulties and 
threshold parameters, demonstrating that students in high active learning classes have significantly higher probabilities of correctly answering the LSCI's items.  
This is further supported by the fact that the average $\Delta \theta$  for high IAS classes is more than twice as large as the average $\Delta \theta$ for low IAS classes.
When we plot 
the average $\Delta \theta$ versus the average normalized gain $\langle g \rangle$ for all classes with at least 25 students, we find a high correlation ($r = 0.99$) between these two 
measures.  A plot of $\Delta \theta$ versus the percentage of class time spent on active learning reproduces the equivalent plot from Prather \emph{et al.} in which $\langle g \rangle$ was 
used as the ordinate variable \cite{prather09}.  
We make the empirical inference from the data that instructors who want their classes to achieve an average improvement in abilities of $\Delta \theta = 1$ logit must spend at 
least 25\% of class time on active learning strategies.  We believe this result supports the idea that faculty who are adopting active learning methods need to do more then simply add a few Peer Instruction or Think-Pair-Share questions every now and then or have students work on problems together in class every couple of weeks.   Instead, using proven active learning strategies needs to become a significant and regular part of their teaching and their formative assessment of learning.  However, the wide range in $\Delta \theta$ for high-IAS classes suggests that just using these strategies often is not enough; one's ability to create an effective classroom environment that incorporates active learning strategies is critical.

Our results also imply that for faculty and STEM education researchers to gain a more complete understanding of the learning of individual students and the effectiveness of a particular class requires more than just a measure of a class's average learning gain, such as $\langle g \rangle$ or $\Delta \theta$.  We plotted the IRT-estimated gain $\Delta \theta$ versus the normalized gain $g$ for all 3205 students in the data set.  Each value of $g$ corresponds to a range of values of $\Delta \theta$.  The size of this range is typically at least 1 logit, which, as noted earlier, represents a significant difference in the probability of giving a correct response to any particular item.  This result suggests that while $\langle g \rangle$ may be good at summarizing the 
average improvement of an entire class, $g$ may not adequately assess individual student learning.

In order to evaluate the effectiveness of different Astro 101 classes represented in the data set, we created histograms of the pre- and post-instruction ability distributions for
each class with at least 25 students.  Such histograms provide information that is not captured by a single number such as $\langle g \rangle$ or average $\Delta \theta$.  An examination of a
class's histogram can reveal to what extent the pre- and post-instruction distributions overlap one another; the smaller the amount of overlap, the greater the fraction of students
in that class who actually experienced a change in their abilities.  Additionally, the histograms reveal how many students are still below the average post-instruction ability, even after a semester of instruction.  In principle, it is possible for a class to have a large average $\Delta \theta$ and still have a majority of its students with below average abilities
post-instruction.  Educators and researchers who are interested in evaluating the overall effectiveness of a class should look at all of these pieces of information in order to 
obtain a more complete understanding of the class.

Item response theory has the power to help researchers and instructors visualize and better understand whether their classes are achieving the kinds of transformative learning experiences they hope to provide for their students.  By sharing the results of IRT analyses with faculty, we have seen them become inspired and empowered to engage in course transformation that they believe can substantially improve the learning experiences for their students.  IRT analyses of student performance, such as the one described in this paper, may be able to play an important role in motivating instructors to adopt active learning methods that have been developed and are supported by research into astronomy and physics education.

\begin{acknowledgments}
This material is based in part upon work supported by the National Science Foundation under Grant No.\ 0715517, a CCLI Phase III Grant for the Collaboration of Astronomy Teaching Scholars (CATS).  Any opinions, findings, and conclusions or recommendations expressed in this material are those of the authors and do not necessarily reflect the views of the National Science Foundation.
\end{acknowledgments}

\appendix
\section{Item Characteristic Curves}
\label{icc}
\begin{figure}
\includegraphics{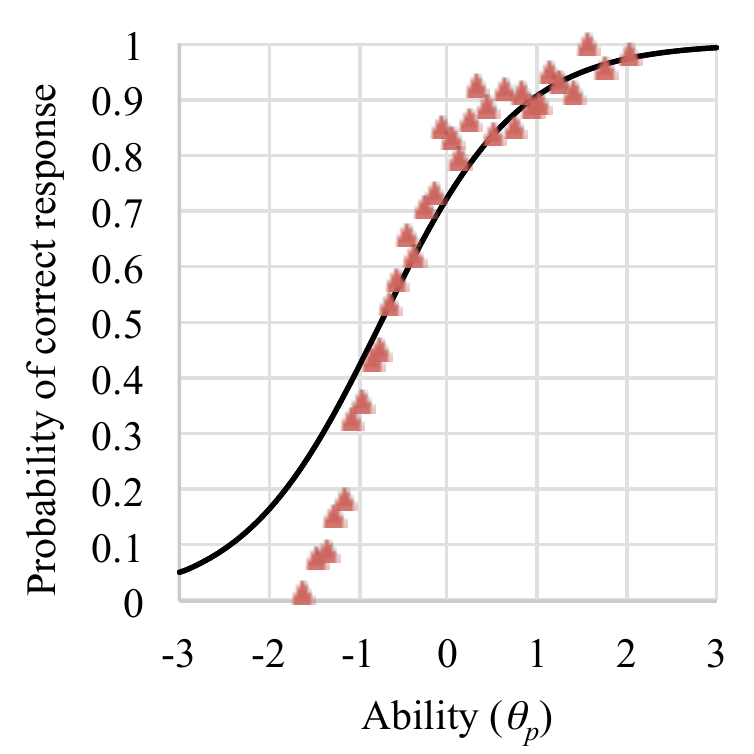}
\caption{\label{icc1}ICC for Item 1.}
\end{figure}

\begin{figure}
\includegraphics{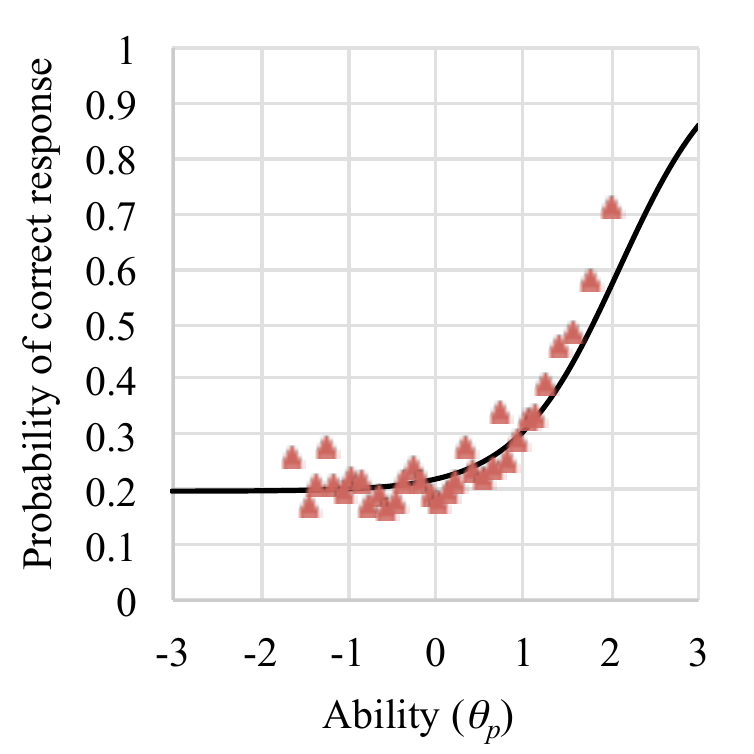}
\caption{\label{icc3}ICC for Item 3.}
\end{figure}

\begin{figure}
\includegraphics{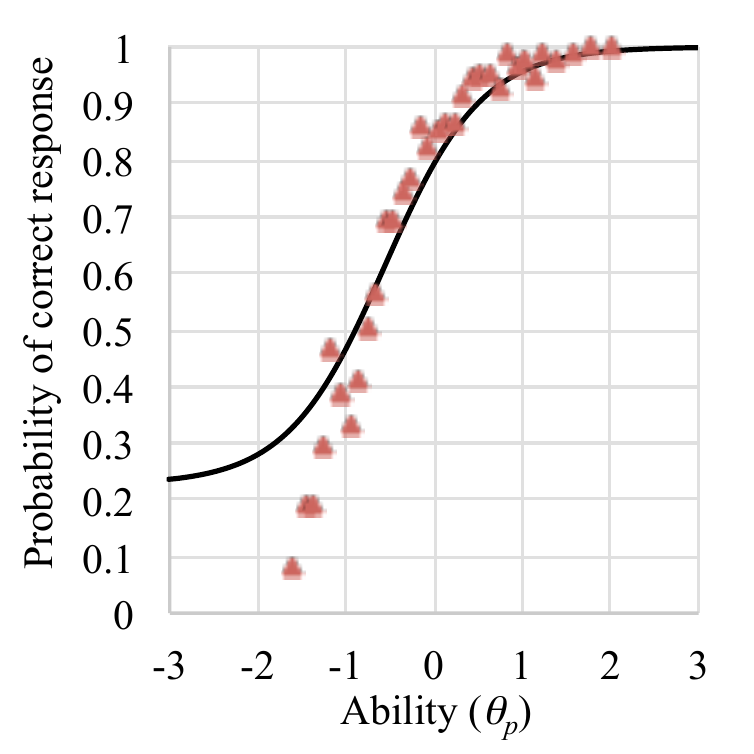}
\caption{\label{icc4}ICC for Item 4.}
\end{figure}

\begin{figure}
\includegraphics{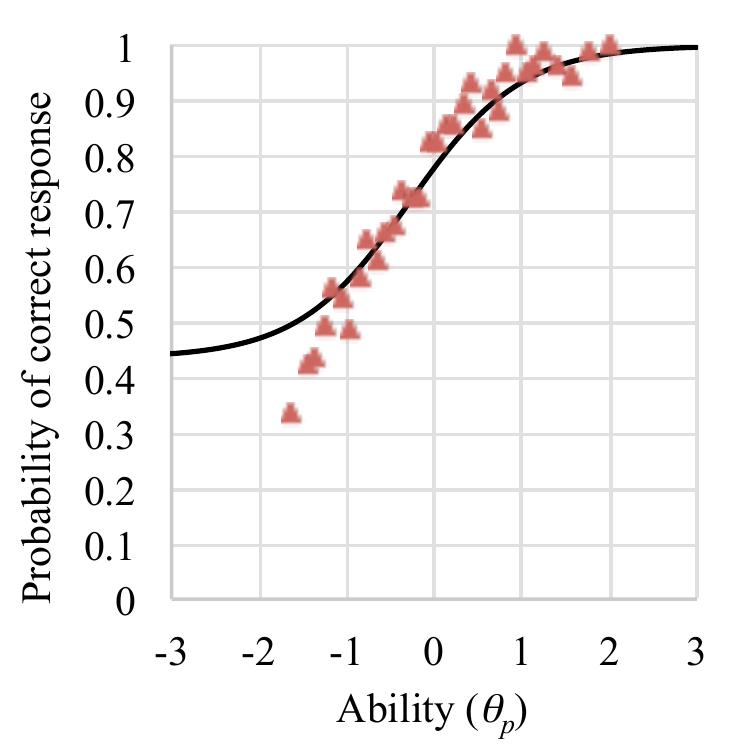}
\caption{\label{icc5}ICC for Item 5.}
\end{figure}

\begin{figure}
\includegraphics{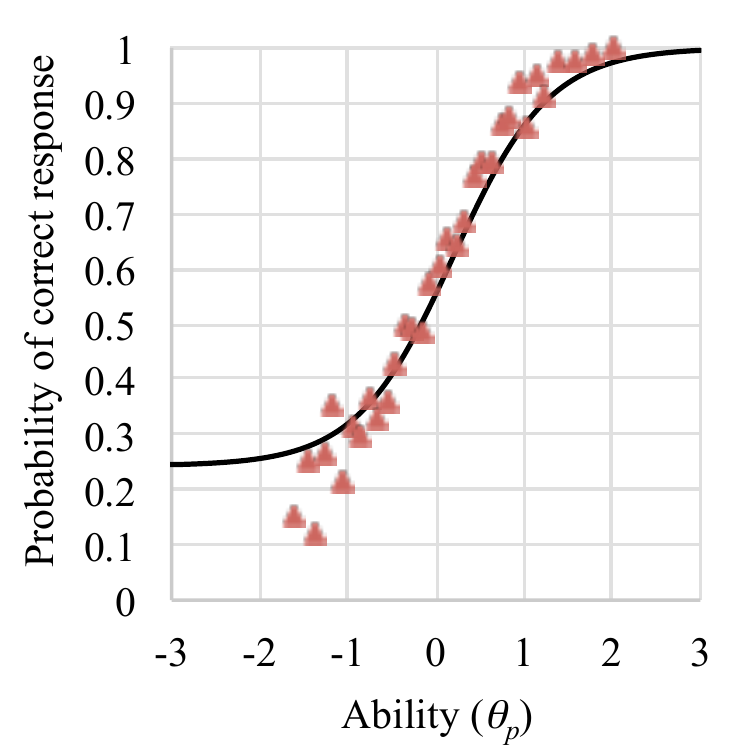}
\caption{\label{icc6}ICC for Item 6.}
\end{figure}

\begin{figure}
\includegraphics{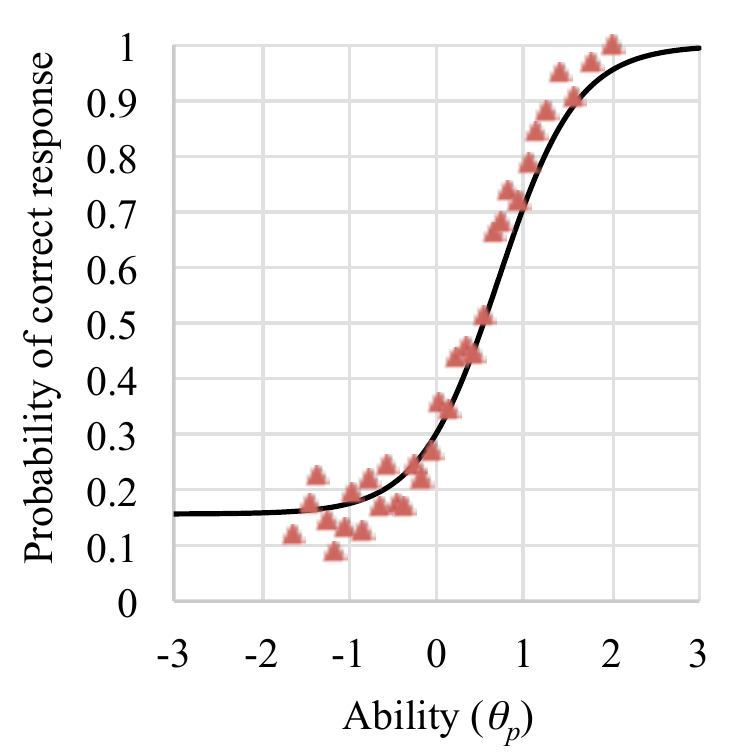}
\caption{\label{icc9}ICC for Item 9.}
\end{figure}

\begin{figure}
\includegraphics{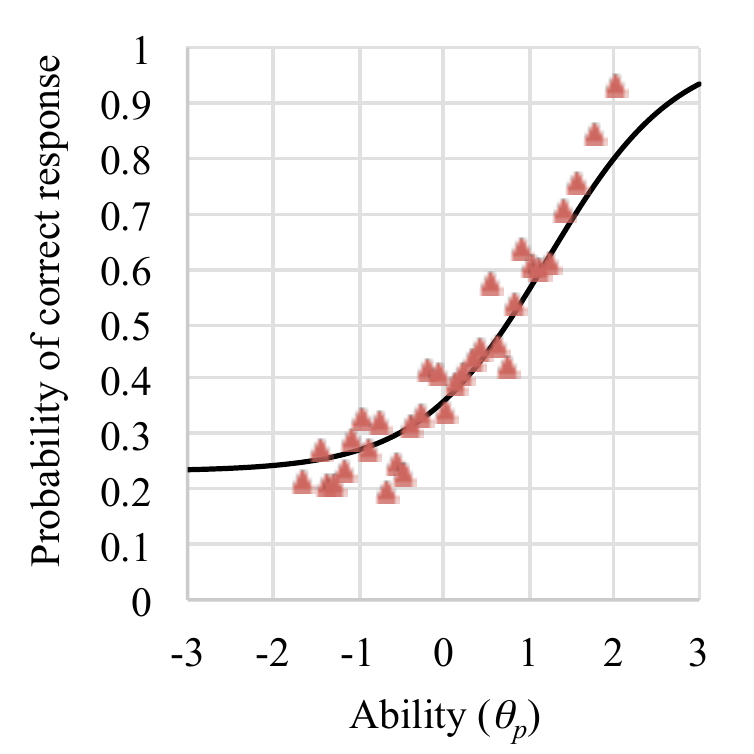}
\caption{\label{icc10}ICC for Item 10.}
\end{figure}

\begin{figure}
\includegraphics{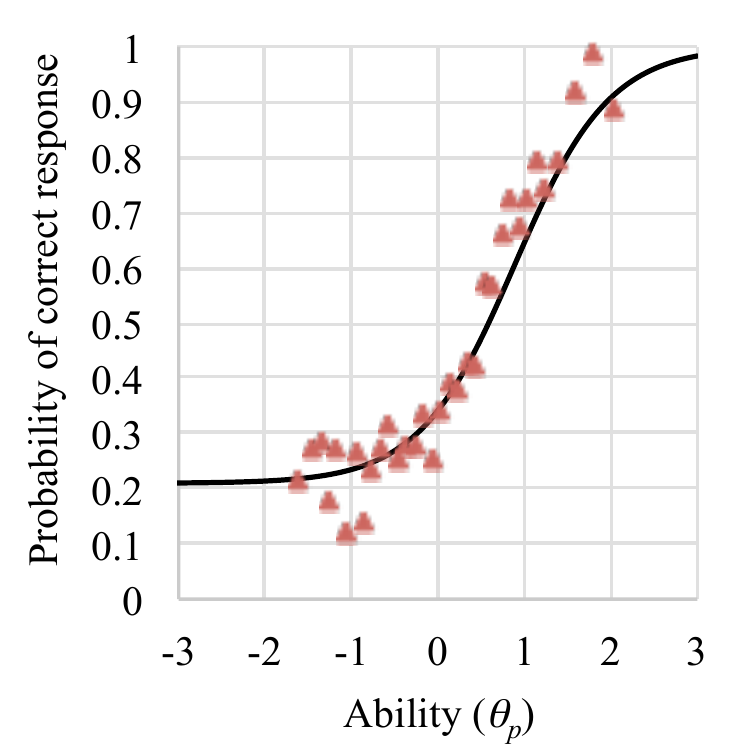}
\caption{\label{icc11}ICC for Item 11.}
\end{figure}

\begin{figure}
\includegraphics{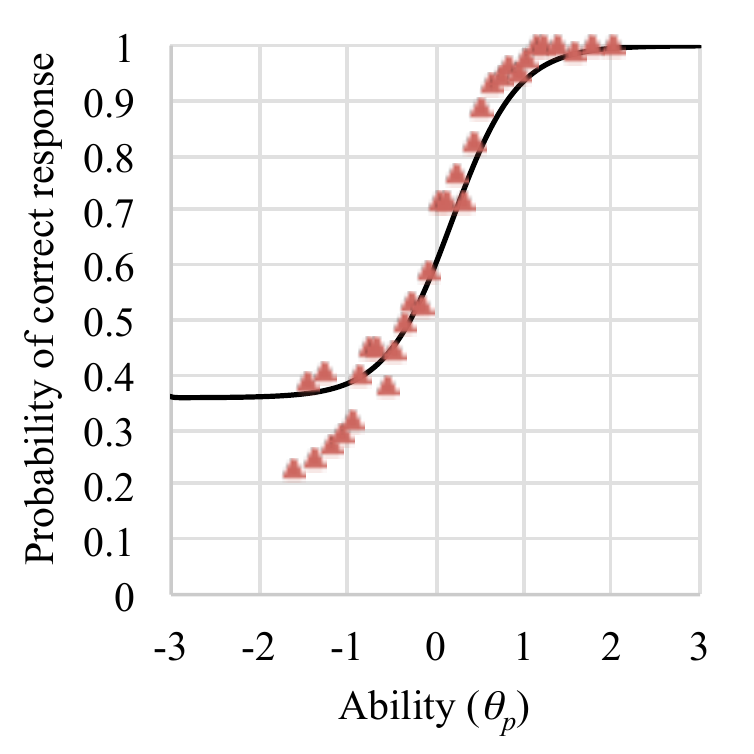}
\caption{\label{icc12}ICC for Item 12.}
\end{figure}

\begin{figure}
\includegraphics{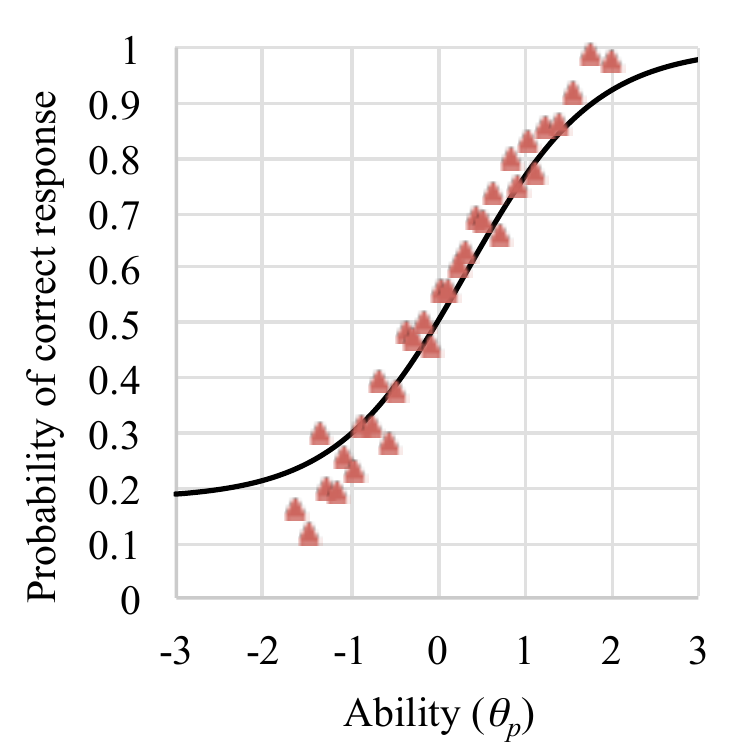}
\caption{\label{icc13}ICC for Item 13.}
\end{figure}

\begin{figure}
\includegraphics{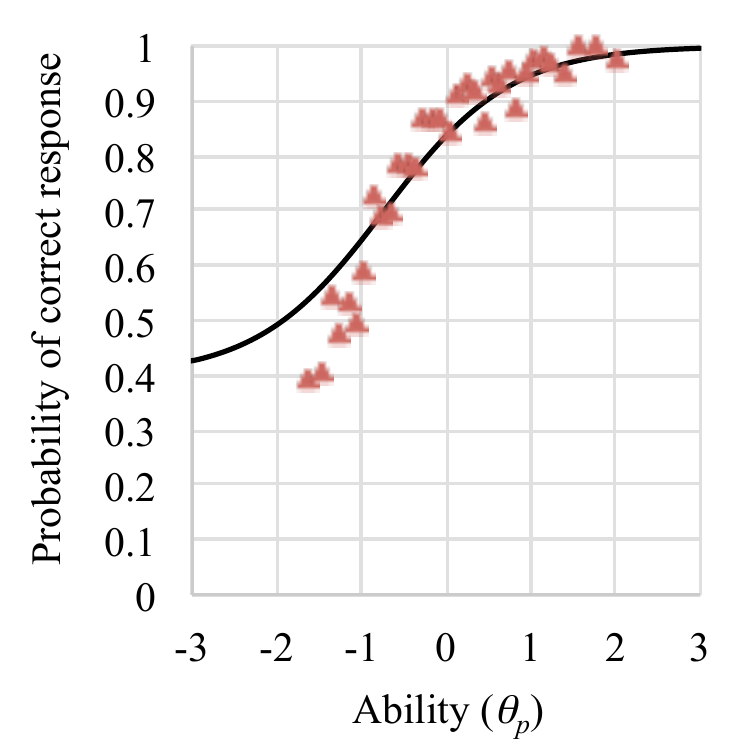}
\caption{\label{icc14}ICC for Item 14.}
\end{figure}

\begin{figure}
\includegraphics{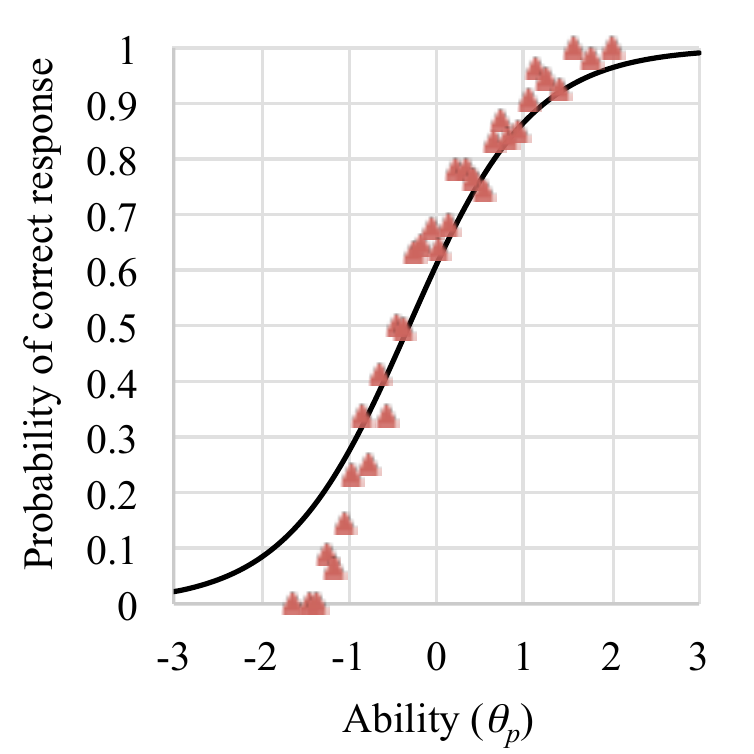}
\caption{\label{icc15}ICC for Item 15.}
\end{figure}

\begin{figure}
\includegraphics{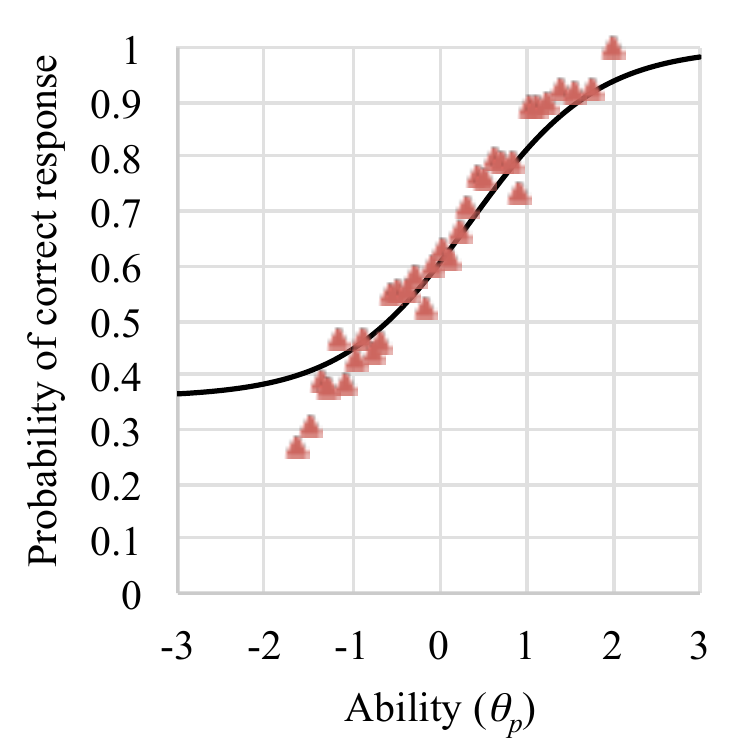}
\caption{\label{icc16}ICC for Item 16.}
\end{figure}

\begin{figure}
\includegraphics{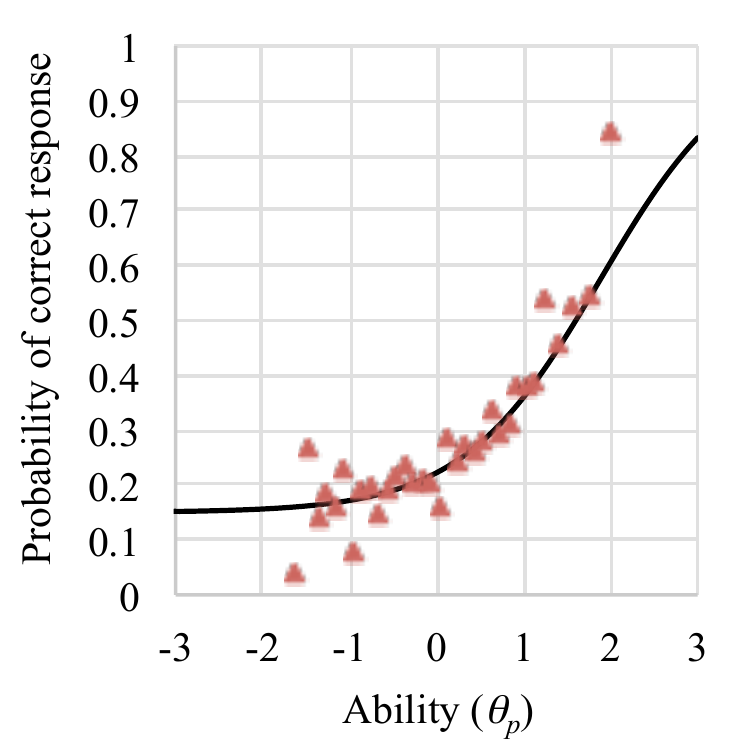}
\caption{\label{icc17}ICC for Item 17.}
\end{figure}

\begin{figure}
\includegraphics{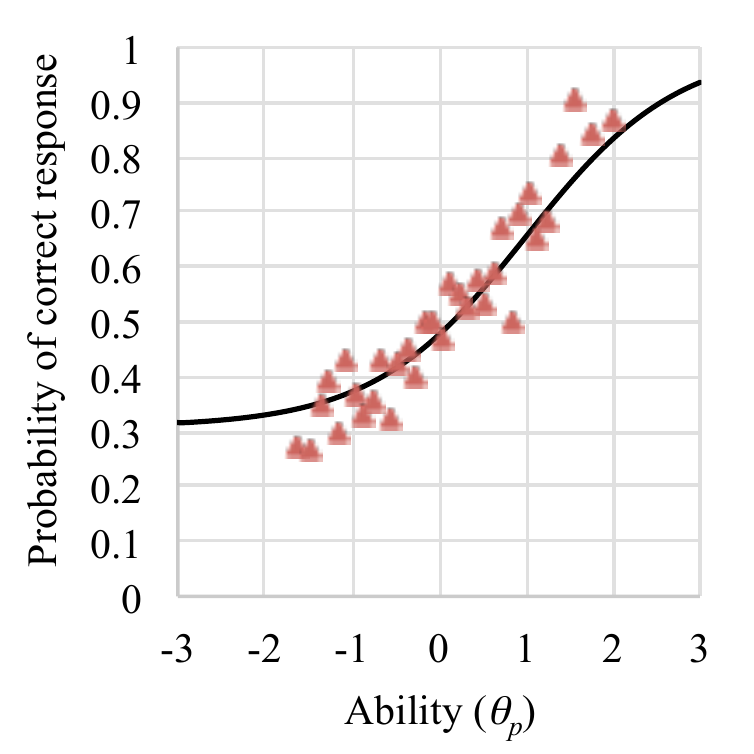}
\caption{\label{icc20}ICC for Item 20.}
\end{figure}

\begin{figure}
\includegraphics{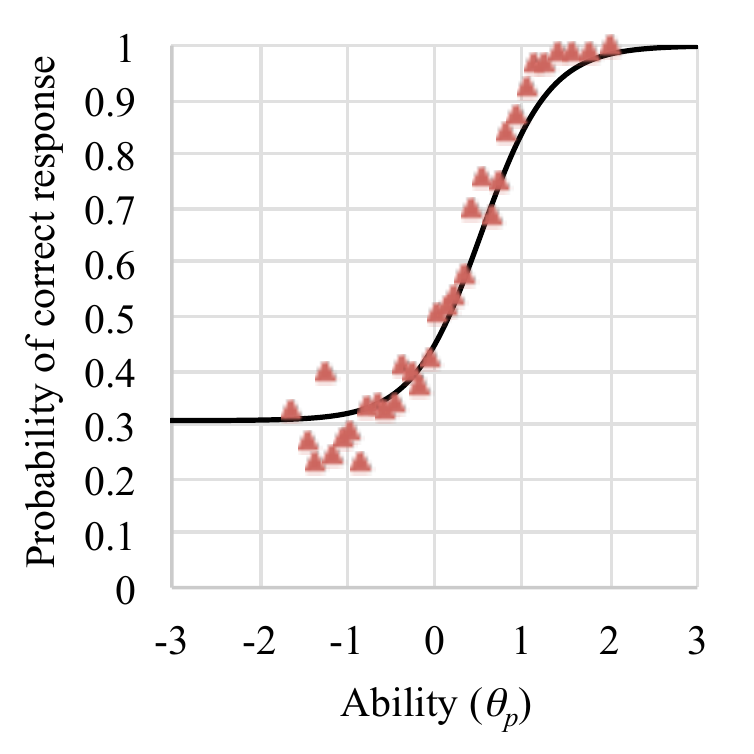}
\caption{\label{icc24}ICC for Item 24.}
\end{figure}

\begin{figure}
\includegraphics{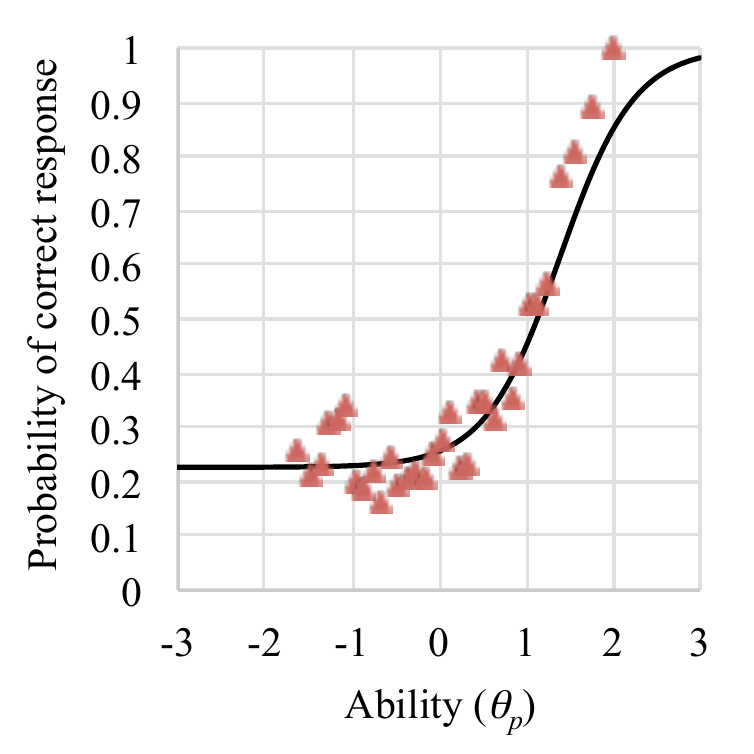}
\caption{\label{icc26}ICC for Item 26.}
\end{figure}

Below are the item characteristic curves (ICCs) for the LSCI's items.  The black curve in Figs.\ \ref{icc1}-\ref{icc26} represents the ICC (the 3PL model-predicted probability of giving a correct response as a function of ability $\theta_p$), while the red triangles represent the average scores of bins of students; these red triangles provide a visual check on the fit of the 3PL model to each item's observed data.  Note that the slope of the ICC is determined 
by the item's discrimination $a_i$, the inflection point is determined by the item's difficulty $b_i$, and the low-ability asymptote is determined by the guessing parameter $c_i$
\cite{embretson00, hambleton93, harris89, wallace10}.

\begin{figure}
\includegraphics{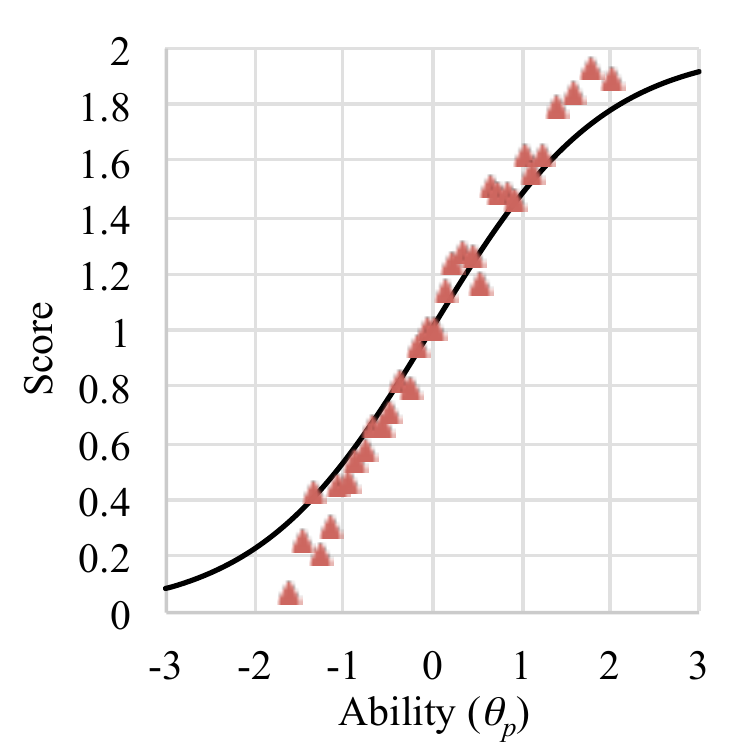}
\caption{\label{icc78}Scores on Item 7 and 8 as a function of ability $\theta_p$.}
\end{figure}

\begin{figure}
\includegraphics{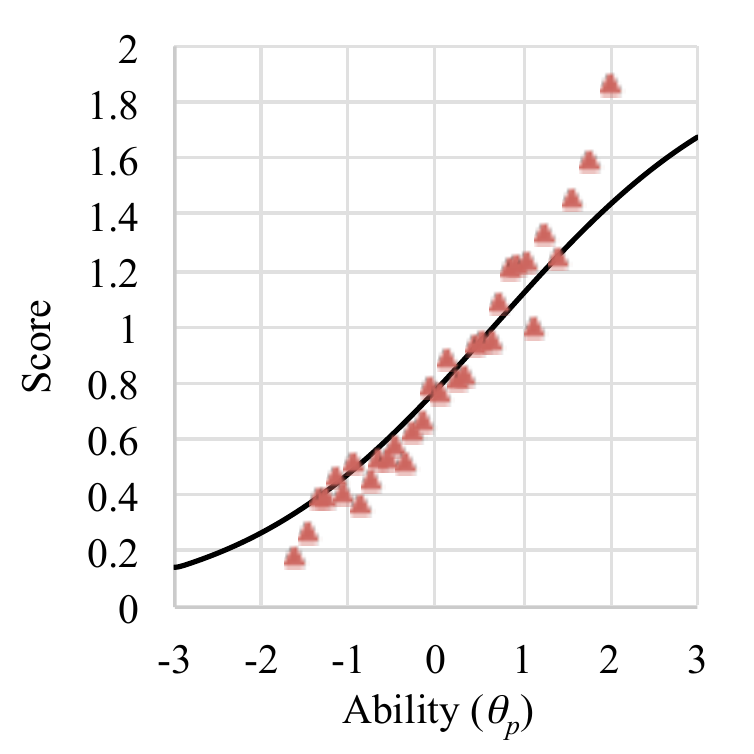}
\caption{\label{icc1819}Scores on Item 18 and 19 as a function of ability $\theta_p$.}
\end{figure}

\begin{figure}
\includegraphics{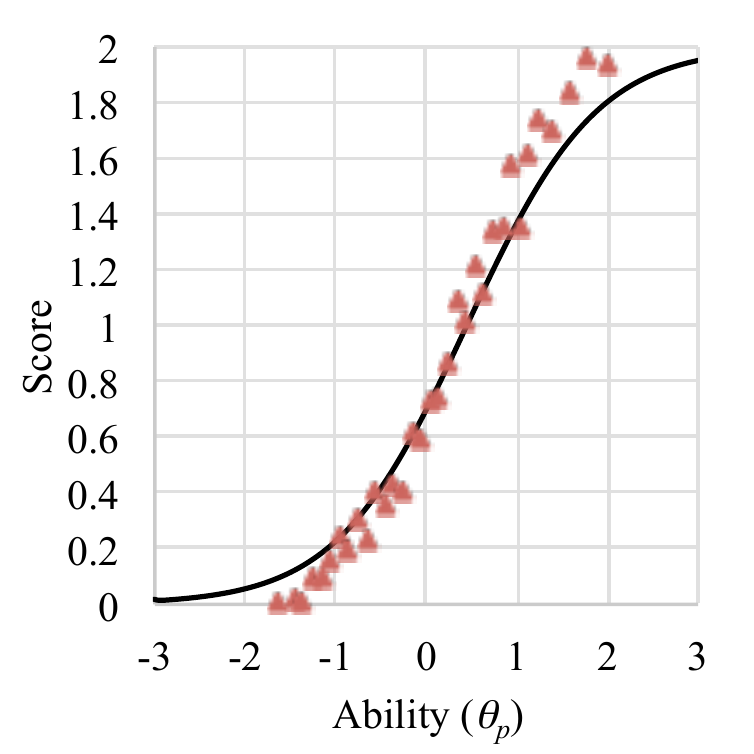}
\caption{\label{icc222}Scores on Item 2 and 22 as a function of ability $\theta_p$.}
\end{figure}

The three polytomous items are represented in Figs.\ \ref{icc78}-\ref{icc222}.  The black curve represents the expected score of a student of ability $\theta_p$ according to
the Graded Response Model.  Expected scores were calculated from the weighted average of the probability of receiving a score of 1 and a score of 2 (i.e., $P(X_{pi} = 1) + 2 P(X_{pi} = 2)$).  The red triangles represent the average scores of bins of students.

\clearpage

\clearpage

\clearpage

\clearpage

\clearpage

\section{Class Histograms}
\label{histograms}
Below are the histograms of pre- and post-instruction ability distributions for all twenty-nine classes in the data set with at least 25 students.  The classes are ordered from high to low average $\Delta \theta$.

\clearpage

\begin{figure}
\includegraphics{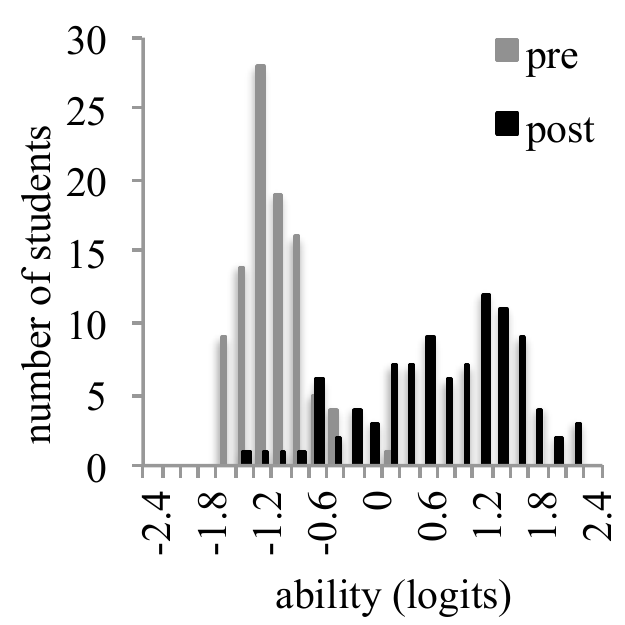}
\caption{\label{class1}The pre- and post-ability histogram for Class 1.}
\end{figure}

\begin{figure}
\includegraphics{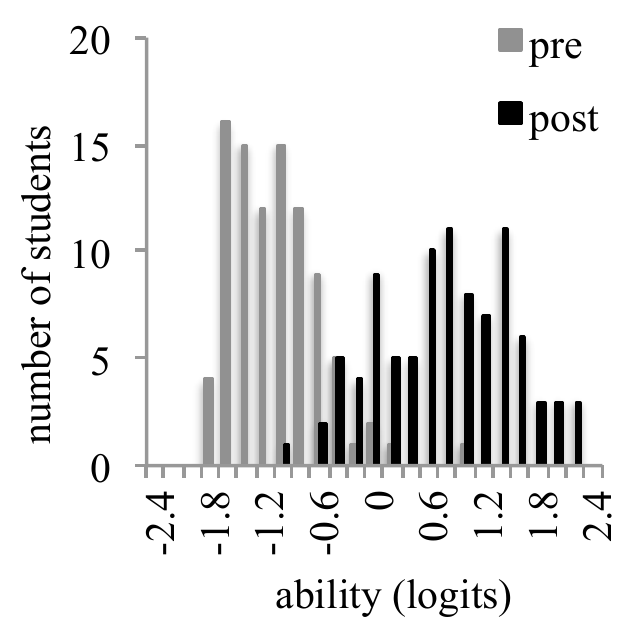}
\caption{\label{class2}The pre- and post-ability histogram for Class 2.}
\end{figure}

\begin{figure}
\includegraphics{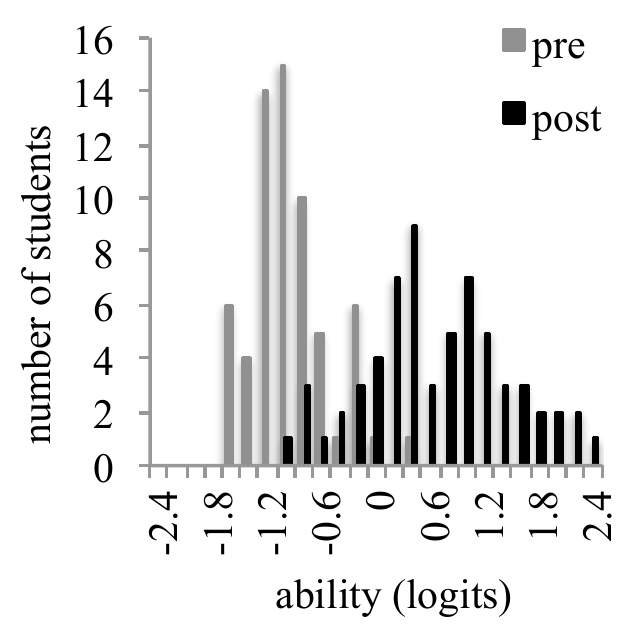}
\caption{\label{class3}The pre- and post-ability histogram for Class 3.}
\end{figure}

\begin{figure}
\includegraphics{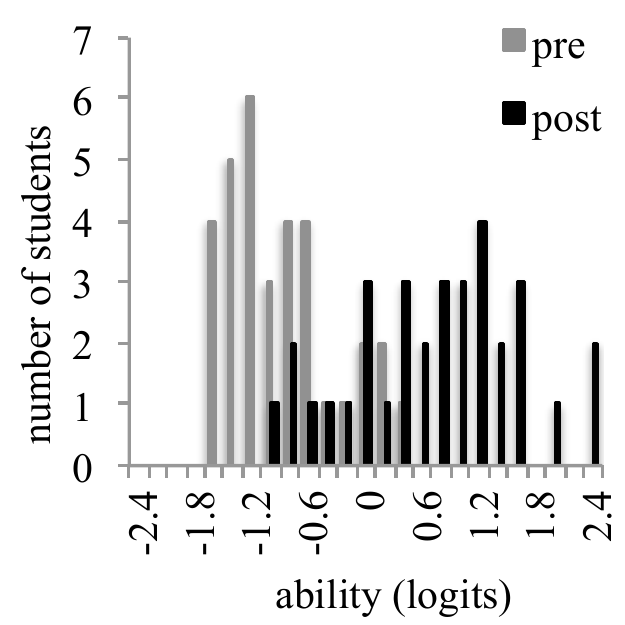}
\caption{\label{class4}The pre- and post-ability histogram for Class 4.}
\end{figure}

\begin{figure}
\includegraphics{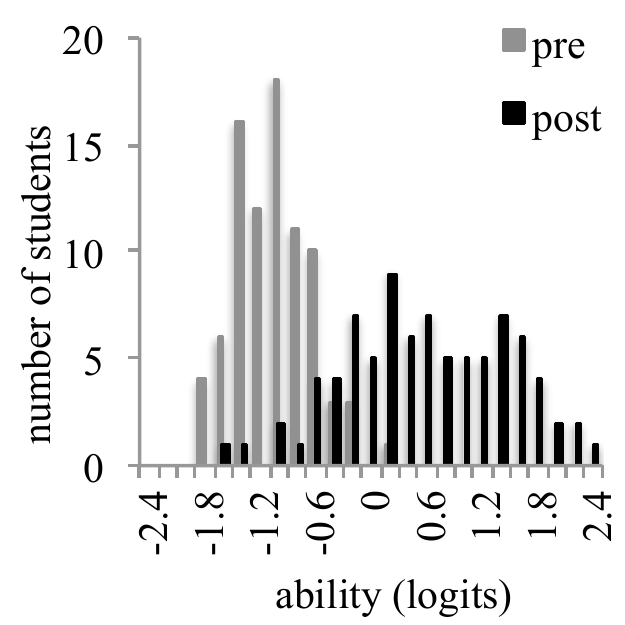}
\caption{\label{class5}The pre- and post-ability histogram for Class 5.}
\end{figure}

\begin{figure}
\includegraphics{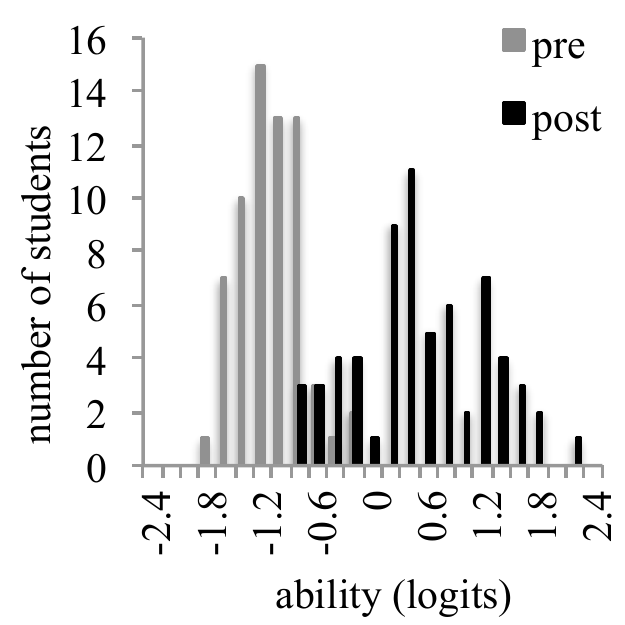}
\caption{\label{class6}The pre- and post-ability histogram for Class 6.}
\end{figure}

\clearpage

\begin{figure}
\includegraphics{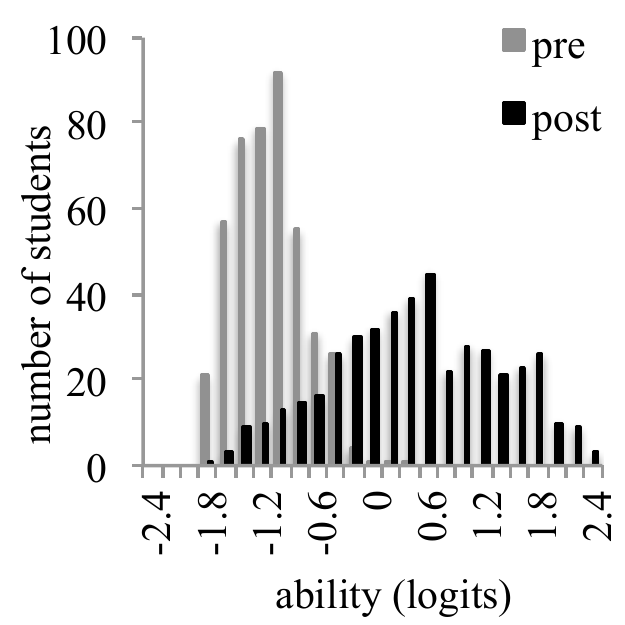}
\caption{\label{class7}The pre- and post-ability histogram for Class 7.}
\end{figure}

\begin{figure}
\includegraphics{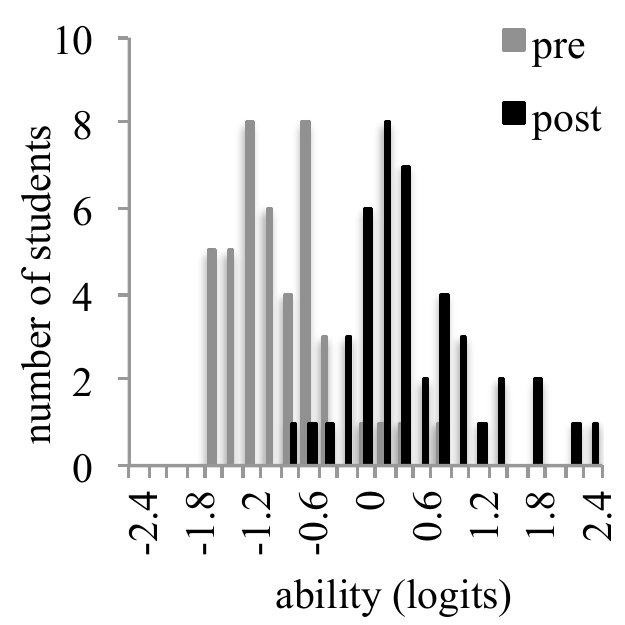}
\caption{\label{class8}The pre- and post-ability histogram for Class 8.}
\end{figure}

\begin{figure}
\includegraphics{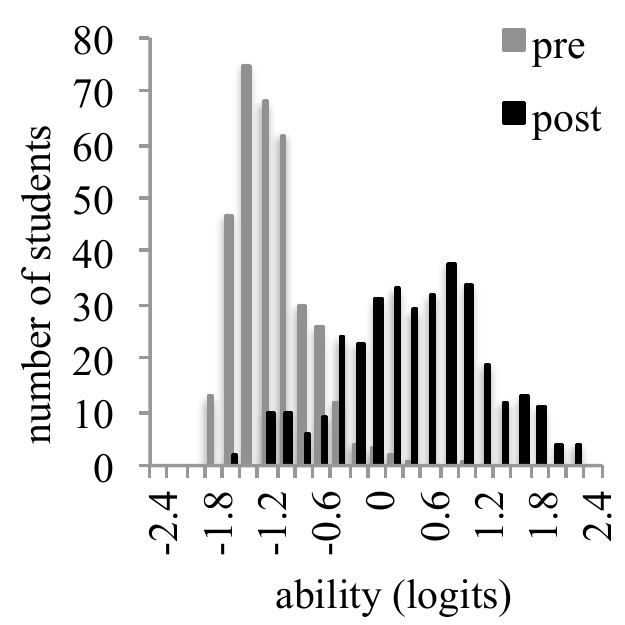}
\caption{\label{class9}The pre- and post-ability histogram for Class 9.}
\end{figure}

\begin{figure}
\includegraphics{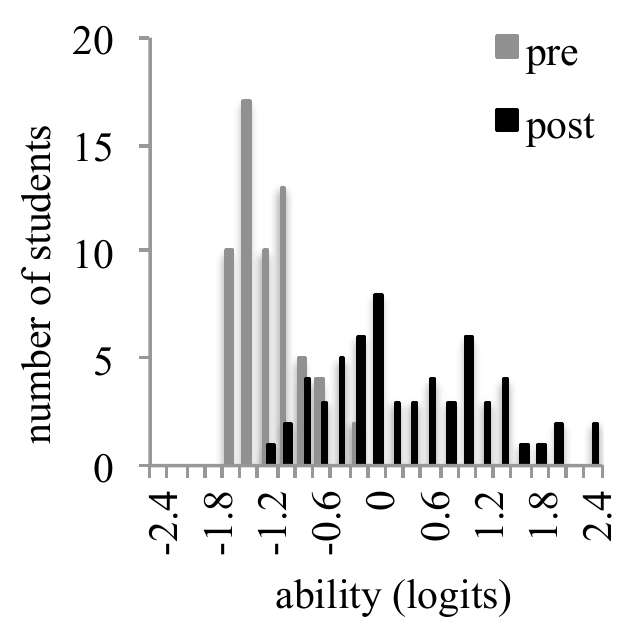}
\caption{\label{class10}The pre- and post-ability histograms for Class 10.}
\end{figure}

\begin{figure}
\includegraphics{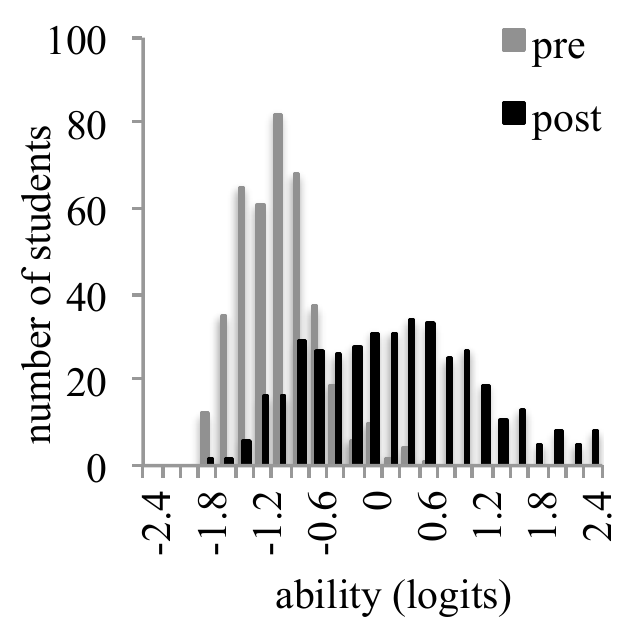}
\caption{\label{class11}The pre- and post-ability histograms for Class 11.}
\end{figure}

\begin{figure}
\includegraphics{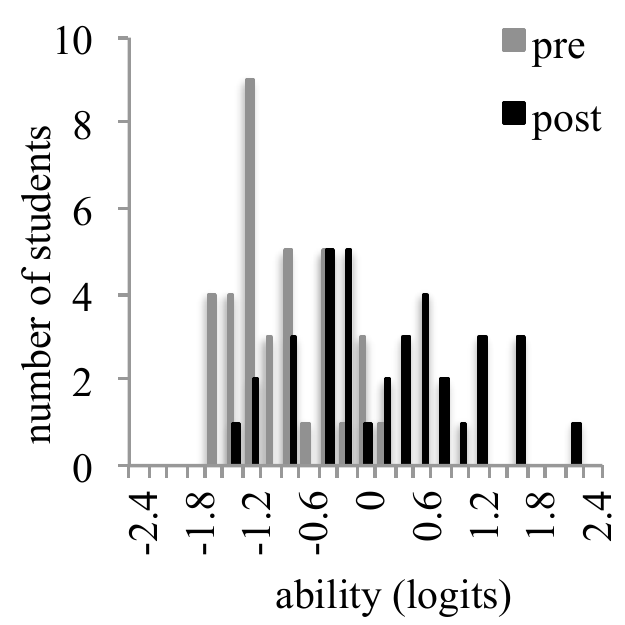}
\caption{\label{class12}The pre- and post-ability histograms for Class 12.}
\end{figure}

\clearpage

\begin{figure}
\includegraphics{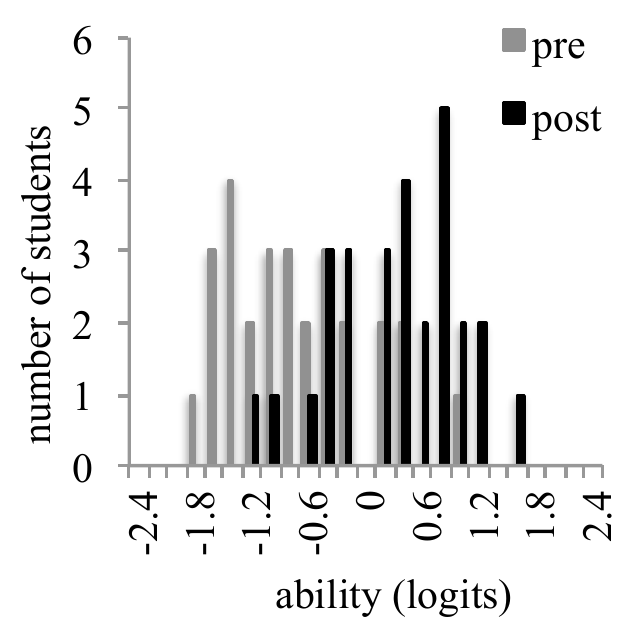}
\caption{\label{class13}The pre- and post-ability histograms for Class 13.}
\end{figure}

\begin{figure}
\includegraphics{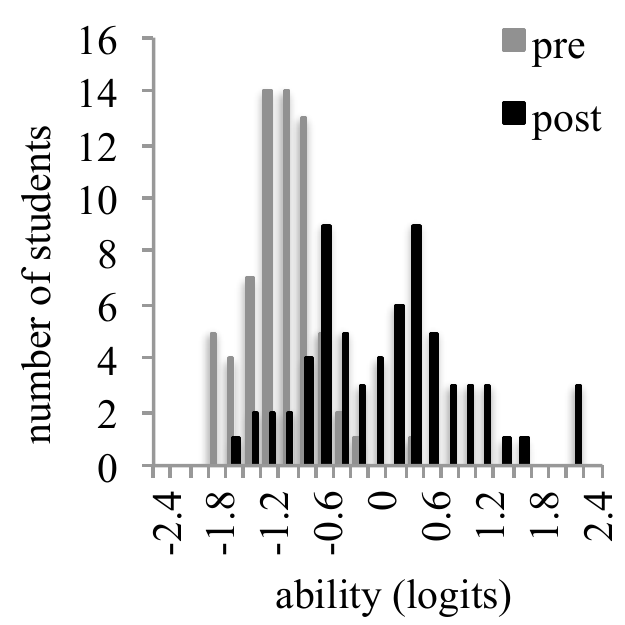}
\caption{\label{class14}The pre- and post-ability histograms for Class 14.}
\end{figure}

\begin{figure}
\includegraphics{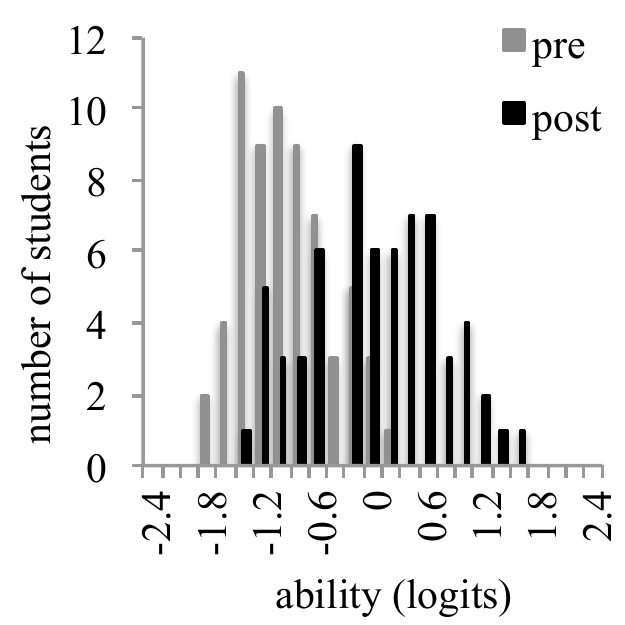}
\caption{\label{class15}The pre- and post-ability histograms for Class 15.}
\end{figure}

\begin{figure}
\includegraphics{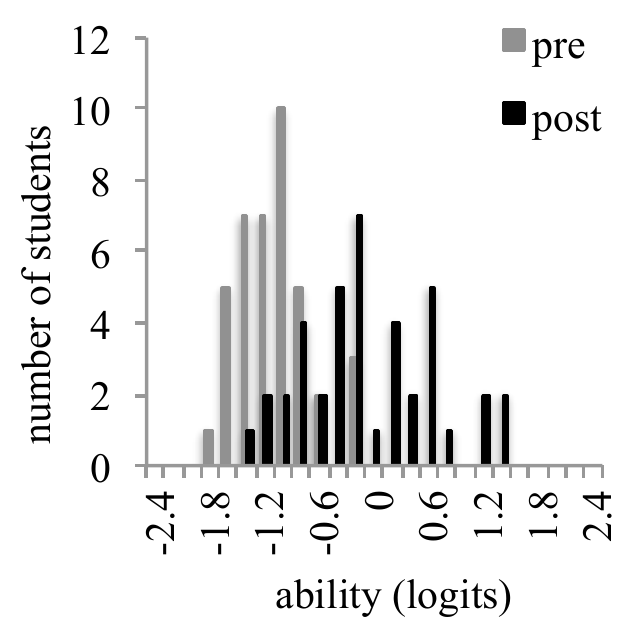}
\caption{\label{class16}The pre- and post-ability histograms for Class 16.}
\end{figure}

\begin{figure}
\includegraphics{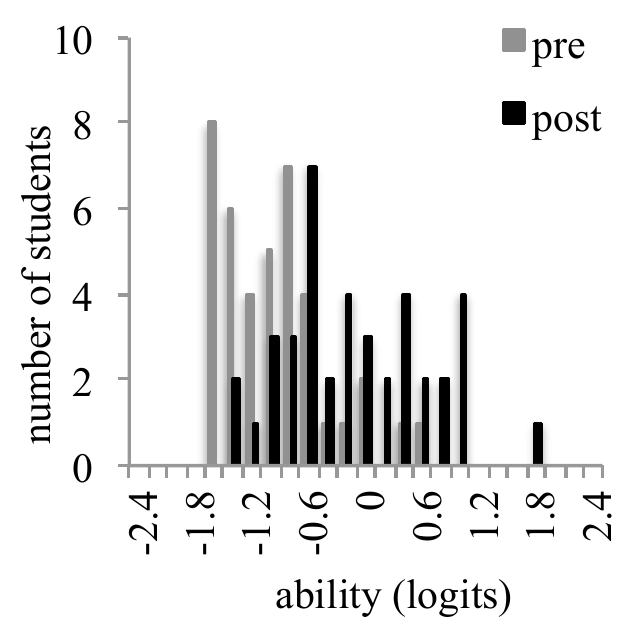}
\caption{\label{class17}The pre- and post-ability histograms for Class 17.}
\end{figure}

\begin{figure}
\includegraphics{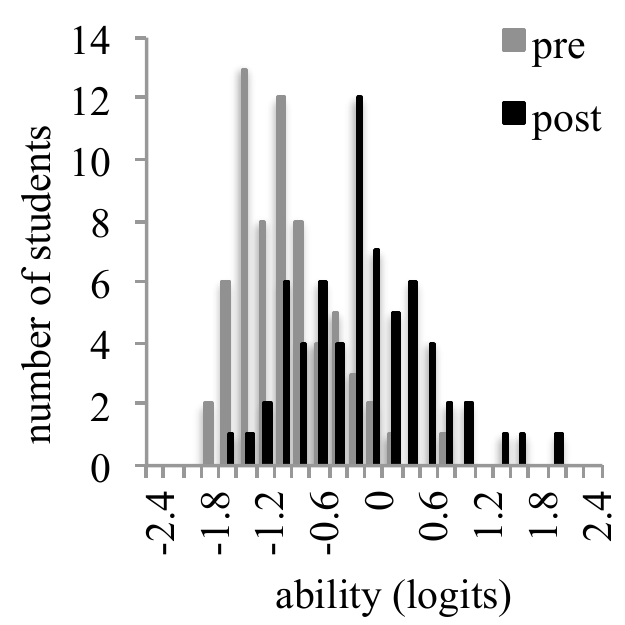}
\caption{\label{class18}The pre- and post-ability histograms for Class 18.}
\end{figure}

\clearpage

\begin{figure}
\includegraphics{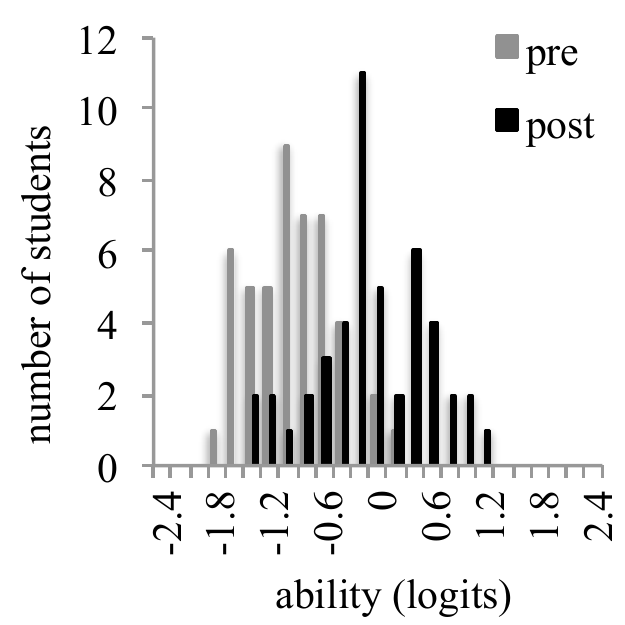}
\caption{\label{class19}The pre- and post-ability histograms for Class 19.}
\end{figure}

\begin{figure}
\includegraphics{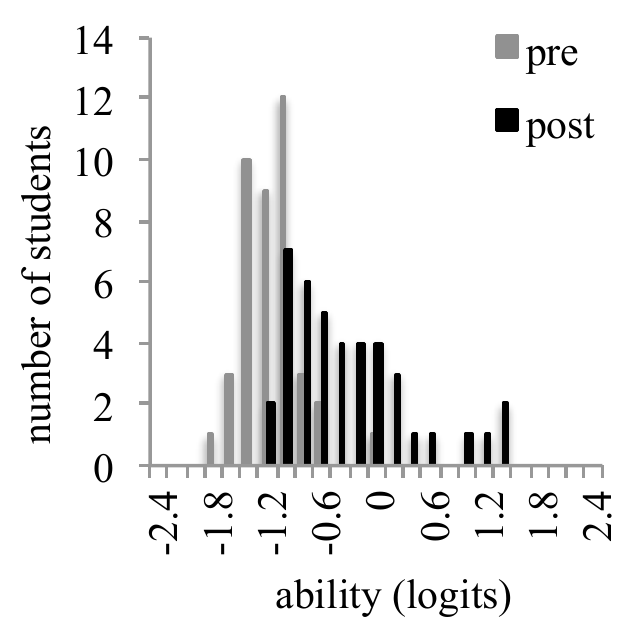}
\caption{\label{class20}The pre- and post-ability histogram for Class 20.}
\end{figure}

\begin{figure}
\includegraphics{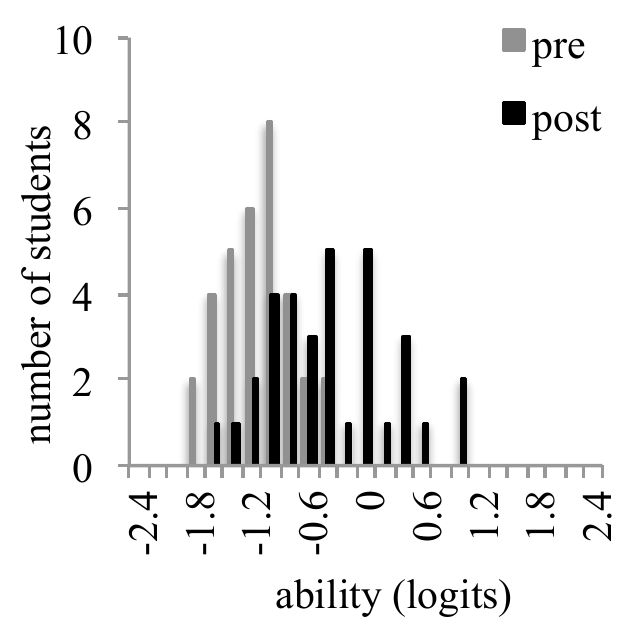}
\caption{\label{class21}The pre- and post-ability histogram for Class 21.}
\end{figure}

\begin{figure}
\includegraphics{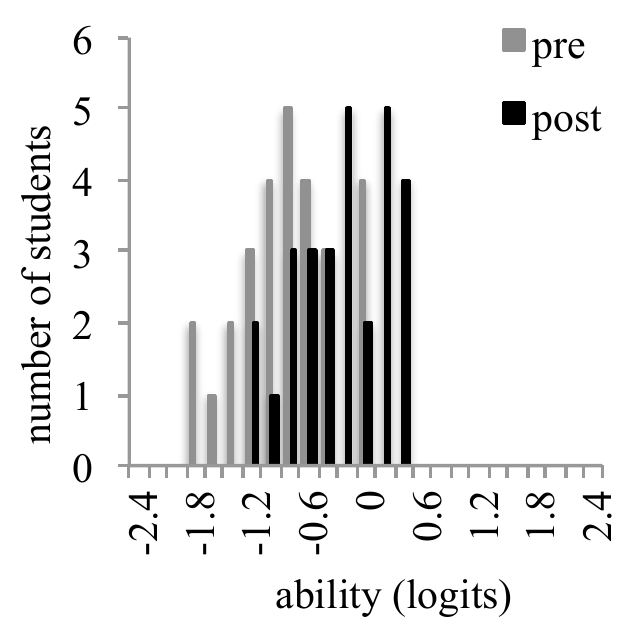}
\caption{\label{class22}The pre- and post-ability histogram for Class 22.}
\end{figure}

\begin{figure}
\includegraphics{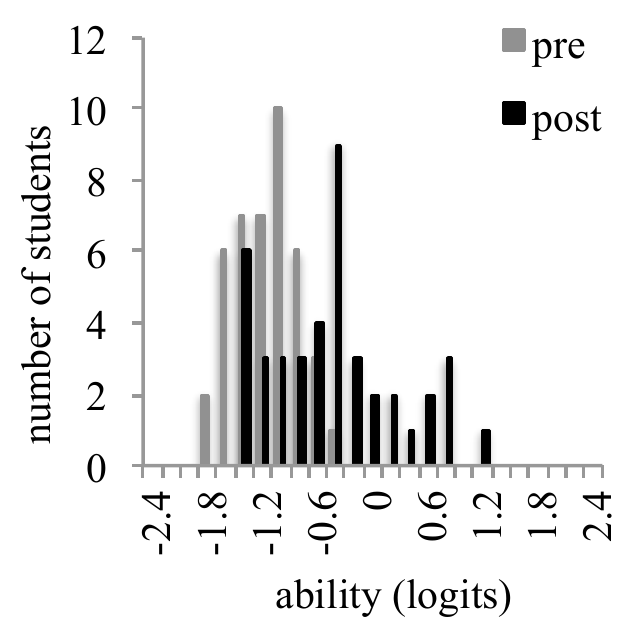}
\caption{\label{class23}The pre- and post-ability histogram for Class 23.}
\end{figure}

\begin{figure}
\includegraphics{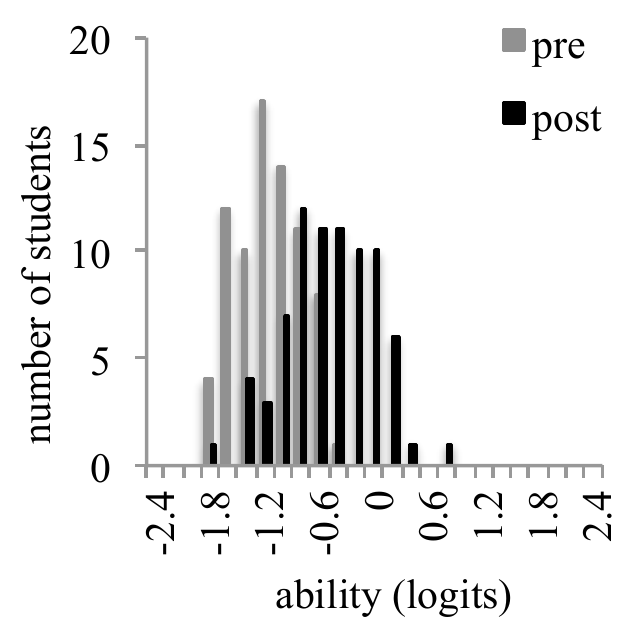}
\caption{\label{class24}The pre- and post-ability histogram for Class 24.}
\end{figure}

\clearpage

\begin{figure}
\includegraphics{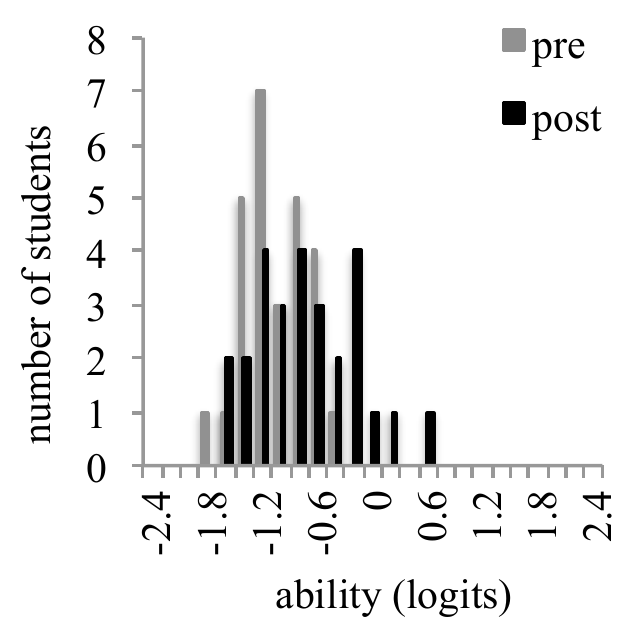}
\caption{\label{class25}The pre- and post-ability histogram for Class 25.}
\end{figure}

\begin{figure}
\includegraphics{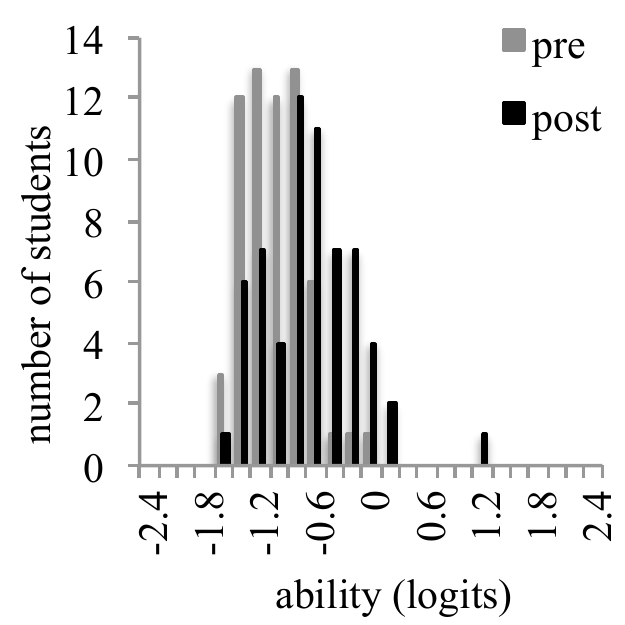}
\caption{\label{class26}The pre- and post-ability histogram for Class 26.}
\end{figure}

\begin{figure}
\includegraphics{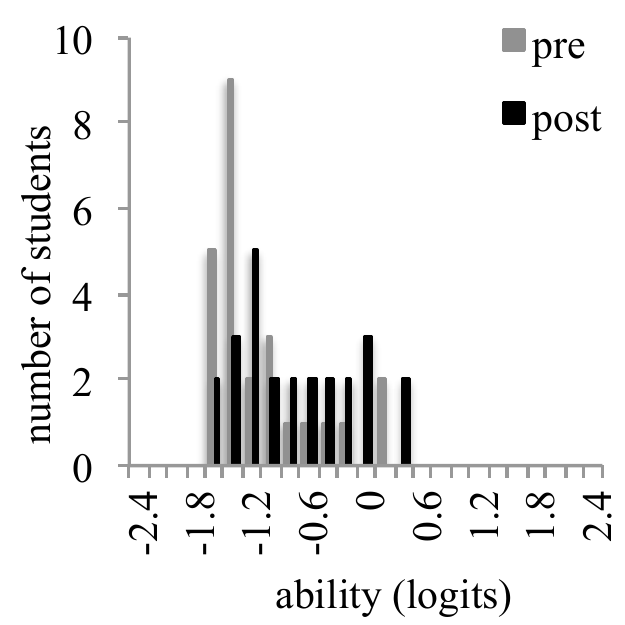}
\caption{\label{class27}The pre- and post-ability histogram for Class 27.}
\end{figure}

\begin{figure}
\includegraphics{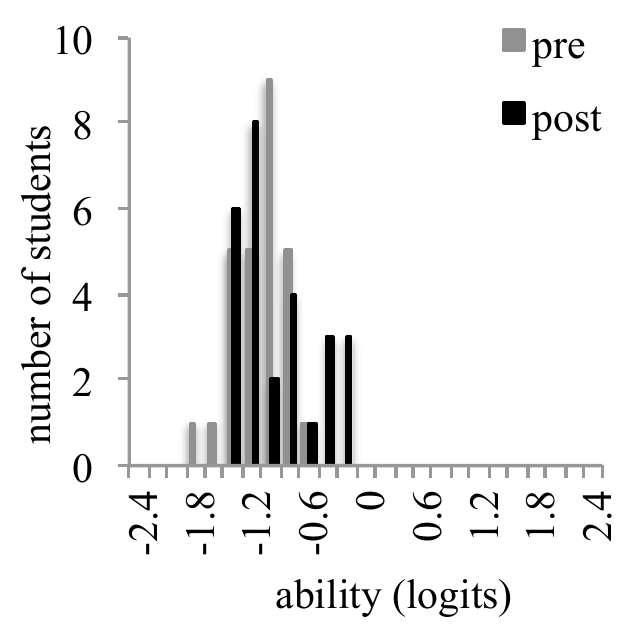}
\caption{\label{class28}The pre- and post-ability histogram for Class 28.}
\end{figure}

\begin{figure}
\includegraphics{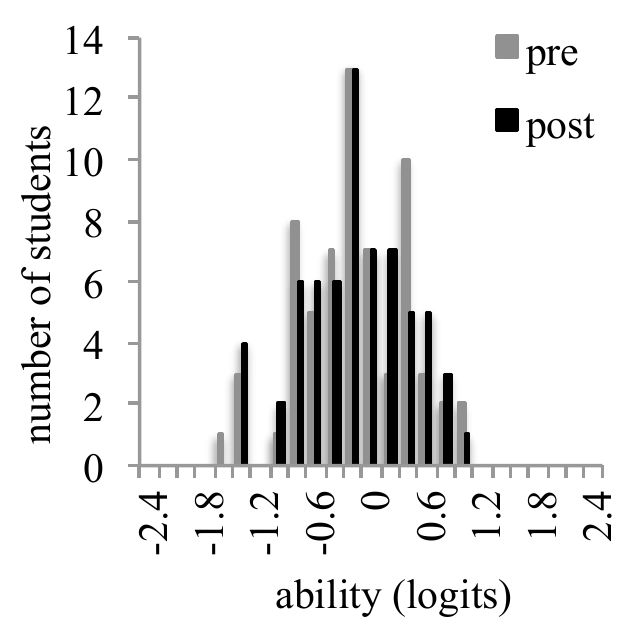}
\caption{\label{class29}The pre- and post-ability histogram for Class 29.}
\end{figure}

\end{document}